\documentclass[twocolumn,aps,pra,superscriptaddress,amsmath,showpacs,tightenlines,pdflatex,longbibliography]{revtex4-2}

\usepackage{amsthm}
\usepackage{xcolor}
\usepackage[normalem]{ulem}
\usepackage{amsmath}
\usepackage{graphicx,times}
 \usepackage{amstext}
\usepackage{amsmath}            
\usepackage{amssymb}            
\usepackage{graphicx}           
\usepackage{latexsym}
\usepackage{blindtext}
\usepackage{bm}
\usepackage{etoolbox}
\usepackage{breqn}
\usepackage{comment}

\makeatletter
\let\cat@comma@active\@empty
\makeatother
\usepackage{url}

\usepackage{url}
\usepackage[colorlinks]{hyperref}
\hypersetup{%
    plainpages=true,
    breaklinks=true,
    hypertexnames=false,
    pageanchor=true,
    colorlinks=true,
    linkcolor={blue},
    citecolor={red},
    urlcolor={blue},
    anchorcolor={black}
    }

\usepackage[position=top,caption=false]{subfig}

\newcommand{\ket}[1]{| #1 \rangle}
\newcommand{\bra}[1]{\langle #1 |}

\newcommand{\beq}{\begin{eqnarray}}
\newcommand{\eeq}{\end{eqnarray}}

\newcommand{\eq}[1]{Eq.~(\ref{#1})}

\theoremstyle{plain}
\newtheorem{theorem}{Theorem}
\newtheorem{lemma}{Lemma}

\def\II{\mathcal{I}}

\def\<{\langle}
\def\>{\rangle}

\def\red#1{{\color{black} #1 \color{black}}}

\def \info#1{}

\def\EB{\text{EB}}

\begin{document}

\title{
Quantifying quantumness of channels without entanglement}
\author{Huan-Yu Ku}
\email{huan\_yu@phys.ncku.edu.tw} \affiliation{Department of
Physics and Center for Quantum Frontiers of Research \& Technology
(QFort), National Cheng Kung University, Tainan 701, Taiwan}
\affiliation{Theoretical Quantum Physics Laboratory, RIKEN Cluster
for Pioneering Research, Wako-shi, Saitama 351-0198, Japan}

\author{ Josef Kadlec }
\affiliation{Palacký University in Olomouc, Faculty of Science, Joint Laboratory of Optics of Palacký University and Institute of Physics AS CR, 17. listopadu 12, 771 46 Olomouc, Czech Republic}

\author{Antonín Černoch}
\affiliation{Institute of Physics of the Czech Academy of
Sciences, Joint Laboratory of Optics of PU and IP AS CR, 17.
listopadu 50A, 772 07 Olomouc, Czech Republic}

\author{Marco Túlio Quintino}
\affiliation{Vienna Center for Quantum Science and Technology (VCQ), Faculty of Physics, University of Vienna, Boltzmanngasse 5,
1090, Vienna, Austria}
\affiliation{Institute for Quantum Optics and Quantum Information (IQOQI), Austrian Academy of Sciences, Boltzmanngasse 3, A-1090
Vienna, Austria}

\author{Wenbin Zhou}
\affiliation{Graduate School of Informatics, Nagoya University,
Chikusa-ku, 464-8601 Nagoya, Japan}

\author{Karel Lemr}
\affiliation{Palacký University in Olomouc, Faculty of Science, Joint Laboratory of Optics of Palacký University and Institute of Physics AS CR, 17. listopadu 12, 771 46 Olomouc, Czech Republic}

\author{Neill Lambert}
\affiliation{Theoretical Quantum Physics Laboratory, RIKEN Cluster
for Pioneering Research, Wako-shi, Saitama 351-0198, Japan}

\author{Adam Miranowicz}
\affiliation{Theoretical Quantum Physics Laboratory, RIKEN
Cluster for Pioneering Research, Wako-shi, Saitama 351-0198,
Japan}
\affiliation{Institute of Spintronics and Quantum Information,
Faculty of Physics, Adam Mickiewicz University, 61-614 Poznań,
Poland} 

\author{Shin-Liang Chen}
\email{shin.liang.chen@phys.ncku.edu.tw} \affiliation{Department
of Physics and Center for Quantum Frontiers of Research \&
Technology (QFort), National Cheng Kung University, Tainan 701,
Taiwan} 
\affiliation{Department of Physics, National Chung Hsing University, Taichung 40227, Taiwan}
\affiliation{Dahlem Center for Complex Quantum Systems,
Freie Universit\"at Berlin, 14195 Berlin, Germany}

\author{Franco Nori}
\affiliation{Theoretical Quantum Physics Laboratory, RIKEN Cluster
for Pioneering Research, Wako-shi, Saitama 351-0198, Japan}
\affiliation{RIKEN Center for Quantum Computing, Wako, Saitama 351-0198, Japan}
\affiliation{Department of Physics, The University of Michigan, Ann Arbor, 48109-1040 Michigan, USA}

\author{Yueh-Nan Chen}
\email{yuehnan@mail.ncku.edu.tw} \affiliation{Department of
Physics and Center for Quantum Frontiers of Research \& Technology
(QFort), National Cheng Kung University, Tainan 701, Taiwan}
\affiliation{Theoretical Quantum Physics Laboratory, RIKEN Cluster
for Pioneering Research, Wako-shi, Saitama 351-0198, Japan}

\date{\today}

\begin{abstract}

Quantum channels breaking entanglement, incompatibility, or
nonlocality are defined as such because they are not useful for entanglement-based, one-sided
device-independent, or device-independent quantum information
processing, respectively. Here, we show that such breaking
channels are related to complementary tests of macrorealism i.e., temporal separability, channel unsteerability, temporal
unsteerability, and the temporal Bell inequality. To demonstrate this we first
define a steerability-breaking channel, which is conceptually
similar to entanglement and nonlocality-breaking channels and
prove that it is identical to an incompatibility-breaking
channel. A hierarchy of quantum non-breaking
channels is derived, akin to the existing hierarchy relations for temporal and
spatial quantum correlations. We then introduce the concept of 
channels that break temporal correlations, explain how they are
related to the standard breaking channels, and prove the following
results: (1) A robustness-based measure for non-entanglement-breaking channels can be probed by temporal nonseparability.
(2) A non-steerability-breaking channel can be quantified by
channel steering. (3) 
Temporal steerability and non-macrorealism can be used for, respectively,
distinguishing unital steerability-breaking channels and nonlocality-breaking channels for a maximally entangled state.
Finally, a two-dimensional depolarizing channel is experimentally
implemented as a proof-of-principle example to demonstrate the hierarchy relation of non-breaking channels using temporal quantum correlations.
\end{abstract}
\maketitle
\section{Introduction}
The extension of quantum physics into the realm of information
theory is important both for fundamental physics and for practical
applications, such as quantum computing, quantum
cryptography~\cite{Gisin2002}, and quantum random number
generation~\cite{Miguel2017,Sanguinetti14}. For the latter
two examples, the practical implementations of entanglement based,
device-independent, and one-side device-independent quantum
information
tasks~\cite{Ekert91,Acin07,Liu2018,Branciard2012} rely
on quantum resources, e.g.,
entangled~\cite{Horodecki09RMP,Ghne2009,Shih-Hsuan2020}, nonlocal~\cite{Einstein35,Bell64,Brunner14RMP,Shih-Hsuan20202,Krivchy2020,Bumer2021},
and steerable states~\cite{UolaRev2020,Wiseman2007,Tan2021,Sarkar2021},
respectively. Extending these ideas to quantum
networks~\cite{Wang2019,Dynes2019,Pompili259,Wehnereaam9288}, one needs reliable quantum
devices (e.g., quantum communication lines~\cite{Cirac1997} and
quantum repeaters~\cite{Briegel1998,Sangouard2011}) to transmit or
generate quantum resources between nodes (senders and receivers)
in the network.

In general, the properties of quantum networks can be
characterized by the concept of quantum
channels~\cite{Chiribella2009}, which is particularly convenient
for estimating the preservability of quantum
resources~\cite{Hsieh2020}. For instance, a reliable quantum
memory~\cite{Langenfeld2020,Xiao2021}
should ideally preserve the entanglement. Therefore, in
channel formalism, the most useful quantum memory is the identity channel, while the threshold of a quantum memory becoming not useful is given by the entanglement-breaking (EB)
channel~\cite{Horodecki2003,Ruskai2003}.

\begin{figure}[tbp]
\includegraphics[width=0.9\columnwidth]{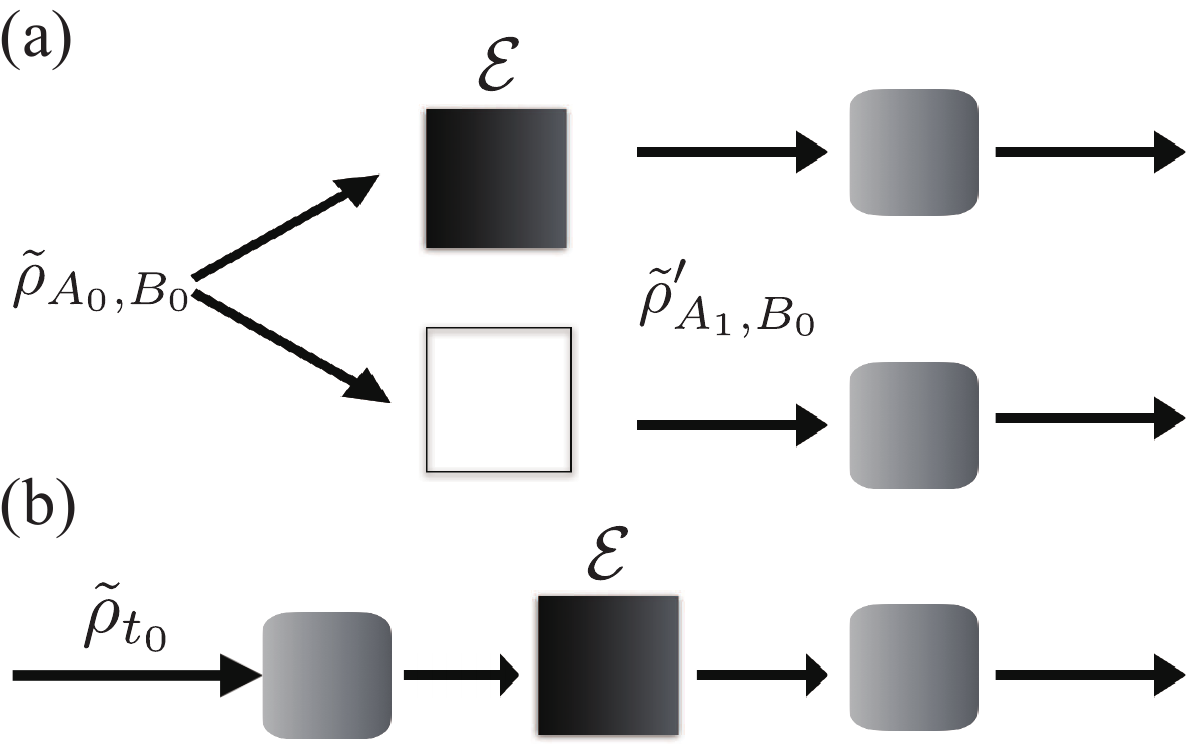}
\caption{Schematic illustration of quantifying  quantumness of channels with (a) spatial and (b) temporal quantum
correlations. In (a), one relies on a bipartite quantum input
$\tilde{\rho}_{A_{0},B_{0}}$ sent to an unknown channel
$\mathcal{E}$ (black box). The channel can be
quantified by analyzing the obtained spatial correlation of the state
$\tilde{\rho}'_{A_{1},B_{0}}$ via a quantum measurement (grey
rectangle). In (b), a given property of $\mathcal{E}$ can be estimated by a temporal quantum correlation with an input
$\tilde{\rho}_{t{_0}}$ which is measured before and after the
channel.}
\label{fig_the_main_idea}\centering %
\end{figure}

\red{In recent years, the framework of a resource theory of quantum memories~\cite{Rosset2018} has been proposed to quantify the quantumness of non-EB channels~\cite{Rosset2018}.
    The experimental quantification~\cite{Graffitti2020} and practical implementation~\cite{Xiao2021} of a quantum memory was demonstrated not long after~\cite{Rosset2018}. These results inspired a new research paradigm around detecting a faithful quantum memory with a set of temporal quantum correlations using  characterized input probing states but uncharacterized measurement apparatus~\cite{Pusey2015} (see Fig.~\ref{fig_the_main_idea} for a schematic view of how to probe a non-breaking channel with temporal quantum correlations).
For instance, when only the sender apparatus is trusted (such a scenario is referred to as the channel-steering scenario~\cite{Piani2015} or a semi-quantum prepare-and-measure scenario~\cite{Guerini2019}), one can certify non-EB channels. 
More recently, sequential-measurement approaches have been proposed to detect quantum memories~\cite{Budroni_2019,Vieira2022temporal} (see also the experimental realization of~\cite{Spee_2020}).
Another approach to witness the non-EB channel is by estimating the coherence of a state sent through the channel~\cite{Simnacher2019}. We emphasize that the methods introduced in these works, and in this article, are different from the typical approach, which used entanglement as a resource to detect quantumness of channels~\cite{Horodecki2003,Ruskai2003}.}

Recently, nonlocality-breaking (NLB) channels~\cite{Pal2015},
defined in a conceptually similar way to EB channels, were shown to be
not useful for device-independent quantum information tasks. As
expected from the hierarchy of correlations~\cite{Jirakova2021}, the EB channel also breaks nonlocality, but not vice
versa~\cite{Pal2015,Heinosaari2015}. Thus, the EB channel is a
strict subset of the set of NLB channels. Although the definition
of the NLB channels is rigorous, one can only assess non-NLB
channels by observing a Bell inequality violation with arbitrary
entangled quantum states as input.

\begin{table*}[!tbp]
\centering
\begin{tabular}{|c|cccc|} \hline \hline
Breaking channels       & EB channel~\cite{Horodecki2003} &
\textit{SB channel}~\cite{Heinosaari2015}& \textit{unital SB
channel}~\cite{Heinosaari2015}& NLB for the MES~\cite{Pal2015}\\ \hline
Spatial correlations & Entanglement& Steerability& Steerability& Bell nonlocality 
\\  \hline
Temporal correlations     & Temporal semi-quantum game*~\cite{Rosset2018} & \textit{Channel steering*}~\cite{Piani2015}& \textit{Temporal steering*}~\cite{Yueh-Nan14}  & \textit{Temporal Bell inequality*}~\cite{Fritz2010,Brukner04}\\
  &\textit{Temporal nonseparability*}~\cite{Fitzsimons2015} & \textit{SQPM}~\cite{Guerini2019} &   & \\
&Coherence~\cite{Simnacher2019}&&&  \\ 
&Sequential measurements~\cite{Budroni_2019,Vieira2022temporal}&&&  \\ \hline\hline
\end{tabular}
\caption{Table showing how to probe non-breaking
channels with spatial and temporal quantum correlations.
Here the
italics represent the results of this work and the asterisk marks
the measures within the corresponding temporal scenarios
satisfying the conditions for a quantum memory monotone. We note that SQPM and MES are the abbreviations of the semi-quantum prepare-and-measure scenario and the maximally entangled state, respectively.
} \label{Table}
\end{table*}

In this work, we propose the concept of a steerability-breaking (SB) channel, which by definition is a channel that is not useful for one-sided device-independent quantum information tasks, and show that it is identical to an incompatibility-breaking channel~\cite{Heinosaari2015}. 
We then introduce a measure for non-SB channels, called the robustness of the non-SB channel. This measure satisfies monotonicity in the sense that the robustness of a non-SB channel cannot increase under non-SB free operations, which map SB channels to SB channels. Similarly, we also propose a measure for non-NLB channels and demonstrate their associated monotonicity. 
We also complete the discussion of the relationship between SB and NLB by
proving that all NLB channels must be SB channels. In
addition, we show that the set of all Clauser-Horne-Shimony-Holt
(CHSH) breaking channels~\cite{Pal2015,Zhang2020}, which is a
particular type of NLB channel, is a strict subset of all SB channels.
Therefore, the hierarchy of breaking channels can be obtained.

\red{We then focus on the quantification of breaking channels using temporal correlations without trust of the input probing states.} 
More specifically, we connect the non-EB, non-SB, and non-NLB channels with
certain temporal quantum correlations including the temporal nonseparability~\cite{Fitzsimons2015}, channel
steerability~\cite{Piani2015,Uola2018}, temporal
steerability~\cite{Yueh-Nan14,Uola2018}, and Leggett-Garg inequalities
(LGIs)~\cite{Leggett85,Emary14} in the form of temporal Bell
inequalities~\cite{Brukner04,Fritz2010,Ringbauer2018,Uola2019,Maity2021,Teh2021Rev}.
Then we show that: (1) temporal nonseparability can be used to measure a quantum memory~\cite{Rosset2018}, (2) channel steering can be used to estimate the robustness of non-SB channels, while the temporal steerability can only quantify non-SB unital channels, and (3) the robustness of non-macrorealism can bound the
robustness of non-NLB channels. In the CHSH scenario, we show that all unital non-CHSH-NLB channels can be certified in the temporal domain. In Table~\ref{Table}, we
summarize some previous observations and our new results about breaking channels and temporal quantum correlations. Finally, we experimentally demonstrate an explicit example to show not only the relationship between breaking channels and temporal quantum correlations, but also the hierarchy relationship between breaking channels.

\section{Quantum correlations and their corresponding breaking channels}
In this section, we first briefly review the definitions and the
properties of the EB and NLB channels. We then propose the SB
channel, which is defined in conceptual analogy to the EB and NLB channels.
The properties of the SB channel, including the relationship with
the incompatibility-breaking channel, will also be discussed. Finally, we
discuss the hierarchy relation for breaking channels.

Before showing our results, we first introduce some notations used
in this work. We consider $\mathcal{L}(H_A)$ as the set of linear operators acting on the Hilbert space $H_A$ with a finite dimension $d_A$. We denote a set of standard density operators
$\mathcal{D}(H_A)\in \mathcal{L}(H_{A})$, satisfying positive
semidefiniteness and having a unit trace.
A quantum channel is described by a set of completely-positive (CP)
trace-preserving (TP) maps from $\mathcal{L}(H_{A})$ to $\mathcal{L}(H_{B})$ as
$\mathcal{O}(H_{A},H_{B})$. Moreover, a unital map is a map that preserves the identity: $\openone\to \openone$.
The set of probability distributions is denoted by
$\mathcal{P(\mathcal{X})}$ with a finite index set $\mathcal{X}$.
Finally, we only consider one subsystem, without loss of
generality, of a bipartite quantum state
$\tilde{\rho}_{A_0,B_0}\in \mathcal{D}(H_{A_0},H_{B_0})$, which is
sent into the quantum channel
$\mathcal{E}\in\mathcal{O}(H_{A_0},H_{A_1})$, and denote the
output state as $\tilde{\rho}'_{A_1,B_0}=(\mathcal{E}\otimes
\openone)\tilde{\rho}_{A_0,B_0}$.

\subsection{Quantum memory and entanglement-breaking channel}
A bipartite quantum state $\tilde{\rho}_{A_0,B_0}$ shared between
Alice and Bob is entangled if the corresponding density operator
is not separable, namely
\begin{equation}
\tilde{\rho}_{A_0,B_0}\neq\sum_{j}p(j)\,\tilde{\sigma}_{{A_0}}(j)\otimes\tilde{\eta}_{{B_0}}(j),
\label{Eq_Sep}
\end{equation}
where $p(j)\in\mathcal{P}(\mathcal{J})$ is a probability
distribution, and $\tilde{\sigma}_{{A_0}}(j)\in
\mathcal{D}(H_{A_0})$ $(\tilde{\eta}_{{B_0}}(j)\in
\mathcal{D}(H_{B_0}))$ is a local density operator. 
In general, the EB channel is defined by sending Alice's subsystem
into a quantum channel $\mathcal{E}\in O(H_{A_0},H_{A_1})$, such
that the entanglement is broken for arbitrary entangled states. We
can explicitly formulate the EB channel as
\begin{equation}
\tilde{\rho}'_{A_1,B_0}=(\mathcal{E}^{\EB}\otimes\openone)(\tilde{\rho}_{A_{0},B_{0}})\in\mathcal{F}_{\text{SEP}}~~\forall~\tilde{\rho}_{{A_0},{B_0}}.
\label{Eq_EB_channel}
\end{equation}
Here, the
superscript $\EB$ is used to denote the channel $\mathcal{E}$ to
be EB, and the set of separable states is denoted by $\mathcal{F}_{\text{SEP}}$.

Entanglement-breaking channels are equivalent to
measure-and-prepare channels, namely
\begin{equation}
\mathcal{E}^{\EB}(\tilde{\rho}_{{t_0}})=\sum_{j}\text{Tr}\left[\tilde{\rho}_{{t_0}}
M_{j}\right]\tilde{\sigma}_{{t_1}}(j), \label{Eq_EB_2}
\end{equation}
where $M_{j}$ is a positive-operator valued measurement (POVM)
element satisfying $M_{j}>0~\forall~j$, and
$\sum_{j}M_{j}=\openone$ with classical outcomes $j$.
Here, with some slight abuse of notation, by $t_0$ ($t_1$) we denote a time indicating that the system is in the Hilbert space  $\mathcal{H}_{t_0}$ ($\mathcal{H}_{t_1}$) before (after) the operation of the quantum channel.
The physical interpretation of the EB channel can be explained as
follows: one measures the original system $\tilde{\rho}_{{t_0}}\in
\mathcal{D}(H_{t_0})$ at $t_0$, after that, based on the outcome $j$, the
corresponding state $\tilde{\sigma}_{{t_1}}(j) \in
\mathcal{D}(H_{t_1})$ is prepared at $t_1$. Obviously, when we send one of
the entangled pairs into a measure-and-prepare channel, the system
becomes separable since one has locally prepared another quantum
state without any direct interaction with the other party.

It has been shown that a non-EB channel is a criterion
for a functional quantum memory because one would like a quantum
memory to, at the very least, preserve the entanglement of a
state. In the framework of the resource theory of quantum
memory~\cite{Rosset2018,Xiao2021}, one can introduce a set of quantum-memory free operations transforming any EB channel to another EB channel (see Appendix.~\ref{App_monotones} for more details).
With these quantum-memory free operations, we can recall that the robustness of a quantum memory is the minimal noise $\alpha$ mixing with the input quantum memory, such that the whole memory is lost, namely
\begin{equation}\label{Eq_R_QM}
\text{R}_{\text{QM}}(\mathcal{E}) = \min  \left\{ \alpha\; \Big|
\frac{\mathcal{E} + \alpha \mathcal{E}'}{1+\alpha}  \in
\mathcal{F}_{\text{EB}} \right\}.
\end{equation}
Here, $\mathcal{E}'$ is an arbitrary quantum channel and  $\mathcal{F}_{\text{EB}}$ is the set of EB channels. It has been shown that the robustness of a quantum memory is a monotonic function after applying a quantum-memory free operation~\cite{Xiao2021}.

\subsection{Nonlocality-breaking channel}
Before introducing the notion of the NLB channel, let us briefly
recall the definition of Bell nonlocality. A spatially separated state is Bell-local when local
measurements with finite inputs $x\in\mathcal{X}$ ($y\in\mathcal{Y}$) and outcomes $a\in\mathcal{A}$ ($b\in\mathcal{B}$) generate a correlation
$p(a,b|x,y)=\text{Tr}\left[(M_{a|x}\otimes
M_{b|y})\tilde{\rho}_{A,B}\right]$ which admits a local-hidden variable (LHV)
model~\cite{Bell64,Brunner14RMP}, namely,
\begin{equation}\label{Eq_HV}
p(a,b|x,y)=\sum_{j}p(j)p(a|x,j)p(b|y,j),
\end{equation}
which the
correlations are pre-determined by a hidden variable $j$. Here, we
denote a set of correlations admitting a LHV model as
$\mathcal{F}_{\text{LHV}}$. Since quantum
correlations are a strict superset of $\mathcal{F}_{\text{LHV}}$ forming a convex set, one can distinguish the local correlation from the quantum ones by testing
the famous Bell inequalities given by the parameters
$\beta^{x,y}_{a,b}$~\cite{CHSH1969,Bell64,Brunner14RMP}, namely
\begin{equation}
B\equiv\sum_{a,b,x,y}\beta^{x,y}_{a,b}\,p(a,b|x,y)\;\leq\;
\delta^{\beta}, \label{Eq_Bell_inequality}
\end{equation}
where $\delta^\beta$ is the local bound for a given Bell
inequality.

Analogous to the EB channels, a NLB channel is the channel under
which a correlation always satisfies a LHV model for arbitrary measurements and
states, namely~\cite{Pal2015}
\begin{equation}
\begin{aligned}
p(a,b|x,y)=&\text{Tr}(M_{a|x}\otimes
M_{b|y}\tilde{\rho}'_{{A_1},{B_0}})\in \mathcal{F}_{\text{LHV}}\\
&\forall~\{M_{a|x}\},\{M_{b|y}\},\tilde{\rho}_{{A_0},{B_0}}.
\end{aligned}
\label{Eq_NLB_channel}
\end{equation}
Similar to the definition of the robustness of a quantum memory, we now consider a noise $\alpha$ mixing with a given channel $\mathcal{E}$, namely
\begin{equation}\label{Eq_R_NLB}
\mathcal{E}_\alpha =  \frac{\mathcal{E} + \alpha \mathcal{E'}}{1+\alpha}.
\end{equation}
This noisy channel always generates a correlation satisfying a LHV model for arbitrary measurements and states. We denote the minimal value $\alpha$ as the robustness of the non-NLB channel $\text{R}_{\text{n-NLB}}(\mathcal{E})$. In Appendix~\ref{App_monotones}, we show that the robustness of a non-NLB channel is a monotonic function under a non-NLB free operation.

NLB channels have some other important known properties, which we summarize here. Unlike the situation with EB channels, the input of a maximally entangled state is not sufficient for verifying if the channel is NLB~\cite{Pal2015,Zhang2019}. 
In particular, because of this, we denote the case with the input being the maximally entangled state $\ket{\Phi}=\sum_i (1/\sqrt{d})  \ket{i}\otimes\ket{i}$ as \emph{NLB channels for the maximally entangled state}. In 
Ref.~\cite{Pal2015}, the authors considered a particular NLB channel [the Clauser-Horne-Shimony-Holt (CHSH) NLB channel], i.e., channels that break CHSH-nonlocality for any state and measurements. Here, we say that a system satisfies CHSH nonlocality if it violates the CHSH inequality, namely:
\begin{equation}
B_{\text{CHSH}}\equiv E(x_1, y_1) +E(x_2, y_1) + E(x_1,y_2) -
E(x_2, y_2)\leq 2, \label{Eq_CHSH_inequality}
\end{equation}
where $2$ is the local bound for the CHSH inequality~\cite{CHSH1969}, $E(x_i,
y_j)\equiv p(a=b|x_i,y_j)-p(a\neq b|x_i,y_j)$ is the expectation
value of $a\cdot b$, for $a\in\mathcal{A}$, $b\in\mathcal{B}$,
$x_i\in \mathcal{X}$, and $y_j\in \mathcal{Y}$, with
$\mathcal{X}=\mathcal{Y}=\{1,2\}$, and $\mathcal{A}=\mathcal{B}=
\{\pm 1\}$. The quantum bound of the CHSH inequality is given by
$2\sqrt{2}$.
We note that the CHSH-NLB channels for the maximally entangled state have the property~\cite{Pal2015}: if the channel is unital, then it is also CHSH-NLB.

\subsection{Steerability-breaking channels}

Now we introduce our first main result: the concept of ``steerability-breaking channel" which breaks any quantum information tasks using quantum steerability as a resource.
We then investigate the properties of the SB channel by showing that the channel is SB if and only if it breaks the quantum steerability of the maximally entangled state. The above property is useful not only for experimental-friendly certifications of SB, but also for the theoretical analysis of SB channels. For instance, (1) we derive that a SB channel is equivalent to the incompatible-breaking channel~\cite{Heinosaari2015}, (2) we propose the robustness as a quantification of a non-SB channel, and (3) we discuss the hierarchy relationship between breaking channels.

Quantum steering refers to the ability of remotely projecting Bob's quantum states by Alice's collection of measurements $\{M_{a|x}\}$ with finite inputs $x$ and outcomes $a$~\cite{Schrodinger35,Wiseman2007}. A set of measurements gives rise to a collection of quantum states, termed as an \emph{assemblage},
\begin{equation}
\rho_{B}(a|x)=\text{Tr}_{A}\left[(M_{a|x}\otimes \openone) \tilde{\rho}_{A,B} \right]
\label{Eq_assemblage}
\end{equation}
where $\tilde{\rho}_{A,B}$ is a bipartite state shared by Alice and Bob. We say that an assemblage is unsteerable when it admits a local-hidden-state (LHS) model, namely
\begin{equation}
\rho_B(a|x)=\sum_{j}p(j)p(a|x,j)\tilde{\rho}_B(j).
\label{Eq_HS}
\end{equation}
Otherwise, the assemblage is steerable.
The physical interpretation of a LHS model is that an assemblage can be pre-determined by a hidden variable $j$, which simultaneously distributes over the statistics $p(a|x,j)$ and the states $\tilde{\rho}_B(j)\in D(H_B)$.
The set of all assemblages admitting a LHS model is denoted as $\mathcal{F}_{\text{LHS}}$.
Violation of a LHS model simultaneously implies that (i) the shared state is entangled and (ii) Alice's measurement violates incompatibility~\cite{Otfried2021IncomRev,Uola2014,Uola2015,Quintino14}. \red{It has been shown that quantum steering is a central resource for one-sided device-independent quantum information tasks including: metrology~\cite{Yadin2021}, quantum advantages on the subchannel-discrimination problems~\cite{Piani2015-2,Sun2018}, key distribution~\cite{Branciard2012}, and random number generation~\cite{Passaro_2015,Skrzypczyk2018,Guo2019}.}

In analogy to the EB and NLB channels, we propose a SB channel as
a channel which breaks the steerability for any collection of
measurements $\{M_{a|x}\}$ acting on the state sent through the
channel $\tilde{\rho}'_{{A_1},{B_0}}$. More specifically, 
the assemblage after a SB channel can always be expressed by a LHS model, namely 
\begin{equation}
\begin{aligned}
\rho_{{B_0}}(a|x)=\text{Tr}_{A_1}&\left[M_{a|x}\otimes\openone \tilde{\rho}'_{{A_1},{B_0}}\right]\in \mathcal{F}_{\text{LHS}}
~\\&\forall~\{M_{a|x}\},\tilde{\rho}_{{A_0},{B_0}}.
\end{aligned}
\label{Eq_SB_channel}
\end{equation}
We denote the set of all SB channels as $\mathcal{F}_{\text{SB}}$.
Moreover, we define $|\mathcal{X}|$-SB channels which break the
steerability with a finite input $|\mathcal{X}|$. For instance, if the finite
index set is $\mathcal{X}=\{1,2,3\}$, we can define the $3$-SB
channel. Here, we show that SB channels have the following properties:
\begin{theorem}\label{theory_SB_maximally_input}
A quantum channel is steerability-breaking if and only if it
breaks steerability of the maximally entangled state.
\end{theorem}
\emph{Proof.---}We present a proof of this theorem in Appendix~\ref{App_SB_IB_are_the_same}. 
We now use the above result to simplify the definition of the SB channel.

The definition of the SB channel is similar to that
of the incompatibility-breaking channel, which maps an
incompatible measurement $\{M_{a|x}\}$ to a jointly measurable one in the Heisenberg picture~\cite{Heinosaari2015,Kiukas17}. More specifically, a set of measurements after incompatibility-breaking channel can always expressed as
\begin{equation}\label{Eq_IB_channel}
\mathcal{E}^{\dagger}(M_{a|x})=\sum_j
p(a|x,j)M_{j}~~\forall~\{M_{a|x}\},
\end{equation}
where $\sum_j p(a|x,j)M_{j}$ is a joint
measurable model with an intrinsic POVM $\{M_{j}\}$ and
postprocessing $p(a|x,j)$. Here, $\mathcal{E}^{\dagger}$ is
the dual map of the quantum channel which is CP and unital. The set of the joint measurements is denoted by $\mathcal{F}_{\text{JM}}$. 
Intuitively, the incompatibility-breaking channels form a proper subset of SB channels because a joint measurement cannot generate a steerable assemblage~\cite{Uola2014,Uola2015,Quintino14}. With Theorem~\ref{theory_SB_maximally_input}, we provide a stronger connection between two breaking channels by the following theorem:
\begin{theorem}\label{theory_IB_SB}
A quantum channel is steerability-breaking if and only if it is incompatibility-breaking.
\end{theorem}
\emph{Proof.---}We present the proof in Appendix~\ref{App_SB_IB_are_the_same}.
We recall that there is a one-to-one relation between an unsteerable assemblage and a joint measurement~\cite{Quintino14,Uola2014,Uola2015}. Our result provides a similar but not identical analog in terms of breaking channels. \red{This result can be naturally extended in a quantitative manner (as shown below).}

To quantify the degree of a non-SB channel,
\red{we consider, again, a noisy channel consisting of a noise $\alpha$ and the input non-SB channel $\mathcal{E}$ [cf. Eq.~\eqref{Eq_R_NLB}]. The standard robustness of the non-SB channel is defined as the minimal value of the noise $\alpha$, such that the whole channel is SB for any measurement set $\{M_{a|x}^{\text{T}}\}$ and any entangled state, namely}
\begin{equation}
\begin{aligned}
    \text{R}_{\text{n-SB}}(\mathcal{E})=\min\{\alpha\big|\text{Tr}_{A_1} [M^{\text{T}}_{a|x}\otimes\openone \mathcal{E}_{\alpha}(\tilde{\rho}_{_{{A_0},{B_0}}})]\in\mathcal{F}_{\text{LHS}}\\ ~\forall\tilde{\rho}_{_{{A_0},{B_0}}},M^{\text{T}}_{a|x}~\},
\end{aligned}\label{Eq:14}
\end{equation}
where $\text{T}$ denotes the transposition.
Inserting the maximally entangled state $\ket{\Phi}=\sum_i (1/\sqrt{d}) \ket{i}\otimes\ket{i}$  (Theorem~\ref{theory_SB_maximally_input}) \red{into Eq.~\eqref{Eq:14}}, we can simplify the standard robustness of the non-SB channel in the Heisenberg picture to arrive at
\begin{equation}\label{Eq_nSBR}
\text{R}_{\text{n-IB}}(\mathcal{E}) = \min  \left\{ \alpha \Big|
\frac{\left[\mathcal{E}^\dagger + \alpha \mathcal{E'}^\dagger\right]}{1+\alpha} (M_{a|x}) \in
\mathcal{F}_{\text{JM}}~\forall~\{M_{a|x}\} \right\}.
\end{equation}
Here, we use the fact that $(\mathbf{X}\otimes \mathbf{Y})\ket{\Phi}\bra{\Phi}=\mathbf{X}\mathbf{Y}^{\text{T}}\otimes \openone\ket{\Phi}\bra{\Phi}$, where $\mathbf{X} (\mathbf{Y})$ is any operator, \red{and the subscript in $\text{R}_{\text{n-IB}}$ is used to denote the non-incompatibility-breaking channels.
One can see that Eq.~\eqref{Eq_nSBR} is the same as the robustness of the non-incompatiability-breaking channel proposed in Ref.~\cite{Guerini2019}.
In Appendix~\ref{App_monotones}, we further show that the robustness of a non-SB channel is a monotonic function under the most general non-SB free operation.}

\begin{figure}[tbp]
\includegraphics[width=1\columnwidth]{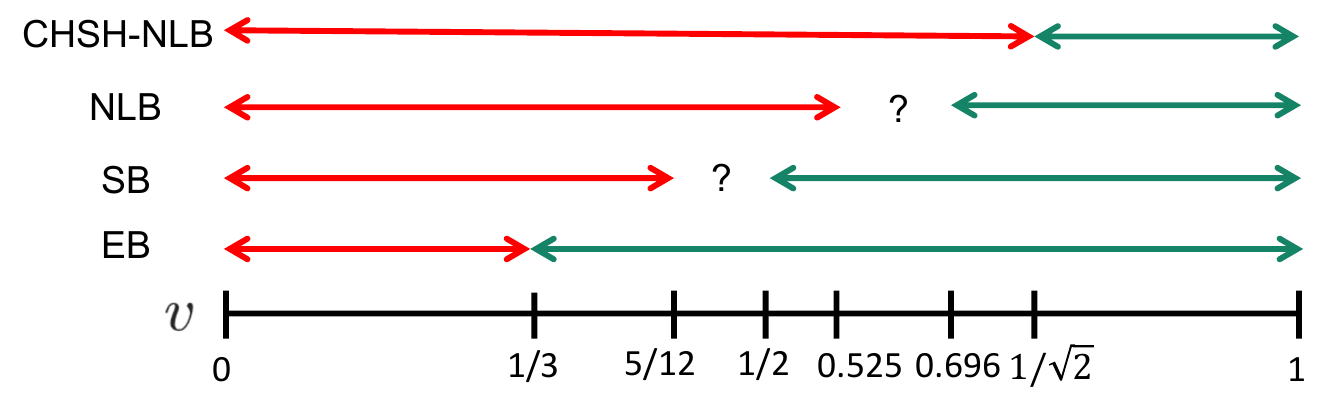}
\caption{Visibility parameter $v$ for which the qubit depolarizing channel is CHSH-nonlocality breaking, nonlocality-breaking (NLB), steerability-breaking (SB), and entanglement-breaking (EB). The first bar in red represents the range, where the channel is proven to break the property, and the second bar in green shows the range, where the channel is proven not to break the property. The white areas with a question mark represent the ranges where the property is not known to be broken.}
\label{fig:breaking}\centering %
\end{figure}

It has been shown that the sets of all incompatibility-breaking
channels (also SB channels) and NLB channels are both supersets of the set of
all EB channels~\cite{Pal2015,Heinosaari2015}. To complete this hierarchy relationship between breaking channels, we show
the following:
\begin{theorem}\label{thm:hierarchy}
The set of all non-EB, non-SB, non-NLB, and non-CHSH-NLB
channels form a strict hierarchy. More specifically, we have the strict inclusions
\begin{equation}
\mathcal{F}_{\text{EB}}\subset\mathcal{F}_{\text{SB}}\subset\mathcal{F}_{\text{NLB}}\subset\mathcal{F}_{\text{NLB}_{\text{CHSH}}}.
\end{equation}
\end{theorem}
\emph{Proof.---}We present the proof of Theorem~\ref{thm:hierarchy} in Appendix~\ref{App_hierarchy}, and we illustrate this theorem for the qubit depolarizing channel
\begin{equation}
\mathcal{E}_{\text{D}}(v,\tilde{\rho})=v\tilde{\rho}+(1-v)\frac{\openone}{2}
\end{equation} 
in Fig.~\ref{fig:breaking}. The experimental demonstration of this theorem is also presented in Sec.~\ref{Sec_exp}. We note that $\mathcal{F}_{\text{NLB}_{\text{CHSH}}}$ denotes the set of CHSH-NLB channels.

\section{Quantifying quantum non-breaking channels with temporal quantum correlations}
In this section, we present our second main result regarding probing the robustness of non-breaking channels with temporal quantum correlations. In section \ref{Certifying EB channel}, we show that
all EB channels are temporally separable and one can probe $\text{R}_{\text{QM}}$ with
temporal nonseparability. In section \ref{Certifying SB channel}, we show that all the non-SB channels can be quantified by
testing channel steerability. In section \ref{Certifying SB and NLB unital channel}, temporal steering and the Leggett-Garg inequality are used to test whether a channel is non-SB and non-NLB, respectively.

Before presenting our results, it is useful to recall the Choi-Jamio{\l}kowski (CJ) isomorphism~\cite{Choi1975285,Jamiokowski1972}, which is a one-to-one mapping between a quantum channel and a positive semidefinite operator. A CJ state $\tilde{\rho}^{\text{CJ}}_{\mathcal{E}}\in
D(H_{t_0}\otimes H_{t_1})$ of a quantum channel
$\mathcal{E}\in O(H_{t_0},H_{t_1})$ is defined as
\begin{equation}
\tilde{\rho}^{\text{CJ}}_{\mathcal{E}}=(\openone\otimes\mathcal{E})\ket{\Phi}\bra{\Phi},
\label{Eq_CJstate}
\end{equation}
where $\ket{\Phi}$ is the maximally entangled state. The output state under the Choi representation is formulated as $\mathcal{E}(\mathbf{X})=\text{Tr}_{t_0}\left[(\openone\otimes \mathbf{X}^{\text{T}})\tilde{\rho}^{\text{CJ}}_{\mathcal{E}}\right]$, with $\mathbf{X}$ being any operator. Note that the quantum state and the quantum channel can be converted with each other with the Choi representation~\cite{Choi1975285,Jamiokowski1972}.
Finally, it has been shown that a CJ state is separable if and only if the corresponding channel is EB~\cite{Horodecki2003,Holevo2008}.

\subsection{Quantifying a non-EB channel with temporal quantum correlations}\label{Certifying EB channel}
We first briefly recall the definition of a pseudo-density operator (PDO), which has primarily been used in the study of the causality of quantum theory~\cite{Fitzsimons2015}. We then show that the temporal nonseparability extracted from a PDO can be used to measure the quality of a quantum memory.
Afterwards, we
show that the PDO formalism is operationally equivalent to the formalism of temporal semi-quantum games in the task of verifying non-EB channels~\cite{Rosset2018}.

To reconstruct the state of a quantum system from trusted measurements performed at different times, one can generalize the concept of a standard density operator in the time
domain~\cite{Fitzsimons2015} (see also Fig.~\ref{Fig_temporal_scenarios}). Without loss of generality, the
events, or observations, collated at different times can be
connected by quantum channels to an input state. In what
follows, we only consider a two-event PDO with a maximally mixed input. The information of a channel is contained in
the PDO, namely~\cite{Pisarczyk2019}
\begin{equation}
P_{\mathcal{E}}=(\openone\otimes\mathcal{E})P_{\mathcal{\openone}},
\label{Eq_PDO}
\end{equation}
where $P_{\mathcal{\openone}}=\text{SWAP}/d$ is the PDO of the
identity channel with the swap operator defined by $\text{SWAP}\ket{ij}=\ket{ji}$. For the qubit case, we have
$\text{SWAP}=1/2\sum_{i=0}^{3}\sigma_i\otimes\sigma_i$. Here,
$\{\sigma_i\}_{i=0,1,2,3}=\{\openone,\hat{X},\hat{Y},\hat{Z}\}$ is a set containing the identity and Pauli operators. 
A PDO is: (1) Hermitian and (2) unit trace but not necessarily
positive semidefinite (the latter is a necessary condition for a
standard density operator). 
Similar to the standard density operator, a PDO is called temporally separable when it admits
\begin{equation}
P_{\mathcal{E}}=\sum_{j} p(j)\,\tilde{\omega}_{{t_0}}(j)\otimes
\tilde{\theta}_{{t_1}}(j), \label{Eq_temporal_separable}
\end{equation}
where $p(j)\in\mathcal{P}(\mathcal{J})$,
$\tilde{\omega}_{{t_0}}(j)\in D(H_{t_0})$, and
$\tilde{\theta}_{{t_1}}(j)\in D(H_{t_1})$. 
In Eq.~\eqref{Eq_temporal_separable}, there exist definite states described by $\tilde{\omega}_{{t_0}}(j)$ and $\tilde{\theta}_{{t_1}}(j)$ at each moment of time, $t_0$ and $t_1$.

It has been shown that the PDO is identical to the partial
transpose of the CJ state of the channel
$\mathcal{E}$~\cite{Jhen-Dong2020,Zhao2018},
\begin{equation}\label{Eq_R=CJ}
P_{\mathcal{E}}^{\text{PT}}=\tilde{\rho}^{\text{CJ}}_{\mathcal{E}},
\end{equation}
where $\text{PT}$ is the partial transposition with respect to the first subsystem. Therefore, a separable PDO implies that the CJ state is also separable and, thus, the corresponding channel must be EB~\cite{Horodecki2003,Holevo2008}. 
With the above properties, we have the following:
\begin{lemma}\label{lemma:PDO and EB}
The partial transposition of the PDO is separable if and only if the corresponding channel is EB.
\end{lemma} 
Recall that one can still use the PDO to distinguish the temporal and spatial correlations, while the partial transposition of the PDO cannot~\cite{Fitzsimons2015}.
To quantify the degree of a quantum memory with a PDO, we can use again the idea of robustness-based measure. Namely, the robustness of a quantum memory is equal to the minimum ratio of a noisy operator one has to mix with the partial transpose of the PDO before the mixture becomes separable. In fact, it can be easily shown that such a measure is the same as the measure derived from the channel formalism in Eq.~\eqref{Eq_R_QM}~\cite{Xiao2021}.

Finally, it is useful to compare the PDO and the temporal
semi-quantum scenario, which certifies all non-EB channels with minimal assumptions in the sense that only state preparation devices are trusted~\cite{Rosset2018}. In detail, the temporal semi-quantum scenario is constructed by an unknown joint measurement, a given quantum channel $\mathcal{E}$, and a set of trusted states. In the beginning, one chooses a quantum state from the set as an input to the quantum channel. A joint measurement is performed on the output state and a state chosen from the same set. One can certify an arbitrary non-EB channel if the set is tomographically complete. From the PDO perspective, we now consider a set of trusted measurements $\{M_{a|x}\}$ at $t_0$ acting on the PDO, which is used for generating a set of states $\tilde{\rho}_{\mathcal{E}}(a|x)\in\mathcal{D}(H_{t_1})=\text{Tr}_{t_0}\left[M_{a|x}\otimes\openone P_{\mathcal{E}}\right]$, up to renormalization, at $t_1$~\cite{Ku2018}. This is the so-called ``normalized" temporal assemblage, and we will formally introduce it later.
The joint measurement is then performed on a characterized quantum input $\{\tilde{\tau}_y\}\in\mathcal{D}(H_{t'_1})$ and the normalized temporal assemblage. The above steps are exactly
the same as the procedure used in temporal semi-quantum
games, and we have
\begin{lemma}\label{lemma:PDO and temporal semi-nonlocal}
The PDO formalism is operationally equivalent to the temporal semi-quantum scenario.
\end{lemma}
\emph{Proof.---}We present a detailed comparison in
Appendix~\ref{App_temporal_semi}. We note that, although in
Ref.~\cite{Zhang2020} the authors have already shown the
relationship between the PDO formalism and the temporal semi-quantum scenario, we provide a clearer physical interpretation in the proof by using the property found in Ref.~\cite{Ku2018}.

\subsection{Certifying non-SB channels with channel steering}\label{Certifying SB channel}
Here, we first recall the concept of channel steering from a PDO perspective. In this way, we can show the  relationship between channel steering and non-SB channels in a quantitative manner. This will allow us to measure all non-SB channels in the temporal domain. We do not consider the more general case of channel steering used in Ref.~\cite{Piani2015} because there it is employed to discuss the coherent properties of an extended channel, which is beyond the scope of this work.

\begin{figure}[htbp!]
\includegraphics[width=1\columnwidth]{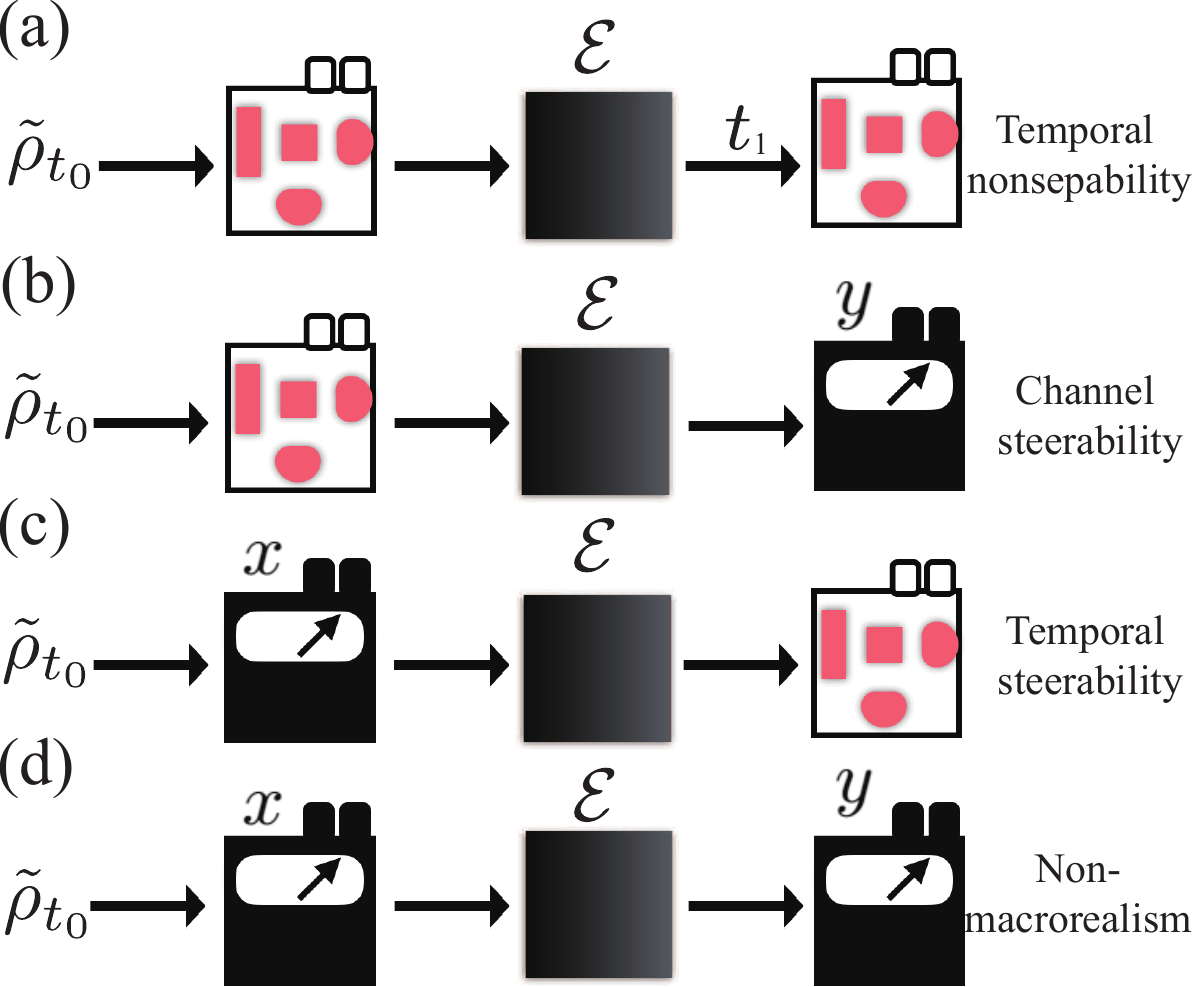}
\caption{Schematic illustration of (a) temporal nonseparability, (b) channel steerability, (c) temporal steerability, and (d) temporal Bell inequality. Here, characterized (uncharacterized) measurements are represented by transparent (black) box. The indices $x\in\mathcal{X}$ ($y\in \mathcal{Y}$) and $a\in\mathcal{A}$ ($b\in \mathcal{B}$) denote the inputs and outcomes of the black box.
}
\label{Fig_temporal_scenarios}\centering %
\end{figure}

In the PDO formalism, the measurements performed at two moments of time are assumed to be characterized. Now, we replace the characterized measurement at time $t_1$ with an uncharacterized one with finite inputs $y\in \mathcal{Y}$ and outcomes $b\in\mathcal{B}$ (see also Fig.~\ref{Fig_temporal_scenarios}). In short, the measurements at times $t_0$ and $t_1$ are trusted and untrusted, respectively. We note that the above scenario is channel steering~\cite{Piani2015}  with only classical outputs (measurement results) at  time $t_1$. To put it another way, the measurements at $t_1$ on the PDO generate a set of evolved states by
\begin{equation}\label{Eq_Channel_steering_PDO}
\text{Tr}_{{t_0}}\left[\openone\otimes M_{b|y}
P_{\mathcal{E}}\right]=\mathcal{E}^{\dagger}(M_{b|y})/d,
\end{equation}
where $d$ is the dimension of the PDO.
Note that the resulting state can be seen as the evolution of the measurement in the Heisenbeg picture.
Since \eq{Eq_Channel_steering_PDO} is a valid assemblage, one can test whether \eq{Eq_Channel_steering_PDO} admits a hidden-state (HS) model, i.e.,
\begin{equation}
\mathcal{E}^{\dagger}(M_{b|y})/d=\sum_j p(j)p(a|x,j)\tilde{\rho}_B(j).
\end{equation}
The above formula suggests that once the HS model is satisfied, the corresponding
measurement set $\{\mathcal{E}^{\dagger}(M_{b|y})\}$ is jointly measurable (see also Appendix~\ref{App_SB_IB_are_the_same}). By Theorem \ref{theory_IB_SB}, if the dual channel breaks the incompatibility of an arbitrary measurement set $\{M_{b|y}\}$, the channel is SB. Therefore, one can use channel steering to certify all SB channels. We also note that if one inserts an EB channel into \eq{Eq_Channel_steering_PDO}, the assemblage also satisfies the HS model. However, due to Theorem~\ref{thm:hierarchy}, not all EB channels can be witnessed with channel steering~\cite{Pusey2015}.

We now define the robustness of channel steering, $\text{R}_{\text{CS}}$, as the minimal ratio $\alpha$ of noise mixed with the dual map of the underlying channel such that the evolved assemblage admits a HS model, or, equivalently, the evolved measurement assemblage admits a jointly measurable model, i.e.,
\begin{equation}\label{Eq_CSR}
\text{R}_{\text{CS}}(\mathcal{E},\{M_{b|y}\}) = \min  \left\{ \alpha \Big|
\frac{\left[\mathcal{E}^\dagger + \alpha \mathcal{E'}^\dagger\right](M_{b|y})}{1+\alpha}  \in
\mathcal{F}_{\text{JM}} \right\}.
\end{equation}
If we now test all incompatible measurements $\{M_{b|y}\}$ in the robustness of channel steering, the minimal one corresponds to the robustness of the non-SB channel in \eq{Eq_nSBR} and we arrive at:
\begin{theorem}\label{theory_channel_steering_IB_channel}
The robustness of the non-steerability-breaking channel or, equivalently, non-incompatible-breaking channels can be quantified by the robustness of channel steering.
\end{theorem}

\subsection{Certifying non-SB and non-NLB channels for maximally entangled states with temporal correlations}\label{Certifying SB and NLB unital channel}
Here, we introduce the last two temporal scenarios: (1) \red{the temporal analogy of quantum steering} (temporal steering) and (2) non-macrorealism in the form of the temporal Bell inequality. Both scenarios can be derived from the assumptions of macrorealism (in the sense that the system is assumed to have well defined pre-existing properties (realism) and can be measured without disturbance (non-invasiveness)~\cite{Emary14}). \red{Beyond quantifying non-Markovianity~\cite{Shin-Liang2016,Bartkiewicz2016,Xiong:17} and  connections to the security of quantum key distribution~\cite{Karol2016}}, we show that temporal steerability can be used to quantify the unital non-SB channels. We also establish for the first time a link between two previously close but not directly related concepts: Bell nonlocality and non-macrorealism. In particular, we show that the value of non-macrorealism provides a lower bound on the magnitude of the non-NLB channel. In other words, if the Leggett-Garg inequality (LGI) can be violated, there exists a corresponding violation of the Bell scenario with the same measurement sets and channel.

In the temporal steering scenario, we consider a set of uncharacterized measurements with finite inputs $x\in\mathcal{X}$ and outcomes $a\in\mathcal{A}$ at $t_0$, while measurements at $t_1$ are assumed to be fully characterized (see Fig.~\ref{Fig_temporal_scenarios}). 
With the above setting, one can obtain a set of subnormalized quantum states, termed a temporal assemblage~\cite{Yueh-Nan14}, namely
\begin{equation}\label{Eq_from_PDO2assemblage}
\rho_{\mathcal{E}}(a|x)=\text{Tr}_{t_0}\left[M_{a|x}\otimes\openone P_{\mathcal{E}}\right].
\end{equation}
We say that a temporal assemblage is temporally unsteerable when it admits an HS model:
\begin{equation}
\rho_{\mathcal{E}}(a|x)=\sum_{j}p(j)p(a|x,j)\tilde{\rho}_B(j)
\end{equation}
[c.f. \eq{Eq_HS}]. In general, an HS model in the temporal domain pre-assigns ontic probability and states in an asymmetric way, such that the system is well defined. For brevity, whenever there is no ambiguity we denote the
assemblage $\{\rho_{\mathcal{E}}(a|x)\}$ as $\vec{\rho}_{\mathcal{E}}$. 
It is convenient to quantify the degree of temporal steerability by the
robustness of temporal steerability~\cite{Ku2016,Maskalaniec2021}, which again refers to the minimum noise $\alpha$ mixed with a given assemblage before the mixture admits an HS model:
\begin{equation}\label{Eq_TSR}
\text{R}_{\text{TS}}(\vec{\rho}_{\mathcal{E}}) = \min  \left\{ \alpha \Big|
\frac{\rho_{\mathcal{E}}(a|x) + \alpha \sigma(a|x)}{1+\alpha}  \in
\mathcal{F}_{\text{HS}} \right\},
\end{equation}
where $\vec{\sigma}$ is an arbitrary assemblage.

From \eq{Eq_from_PDO2assemblage} and the definition of the PDO, one
immediately sees that the temporal assemblage
$\vec{\rho}_{\mathcal{E}}$ can be formulated as
\begin{equation}
\rho_{\mathcal{E}}(a|x)=\mathcal{E}(M_{a|x}) /d.
\end{equation} 
Because now the quantum channel is acting on the measurement, in
what follows we only consider the unital quantum channel denoted as $\mathcal{E}_{\text{unital}}$. It is easy to see that the set of quantum channels is a superset of the unital quantum channels. Obviously, if
$\text{R}_{\text{TS}}(\vec{\rho}_{\mathcal{E_{\text{unital}}}})=0$ when
considering an arbitrary measurement set $\{M_{a|x}\}$, then the
channel $\mathcal{E}_{\text{unital}}$ is a unital SB channel. With Theorem~\ref{theory_SB_maximally_input}, we can certify all unital
SB channels with temporal steering. Furthermore, $\text{R}_{\text{TS}}(\vec{\rho}_{\mathcal{E}_{\text{unital}}})$ provides a lower bound on the $\text{R}_{\text{n-SB}}(\mathcal{E}_{\text{unital}})$ in \eq{Eq_nSBR}. If the entire measurement set $\{M_{a|x}\}$ is considered, we have $\text{R}_{\text{TS}}(\vec{\rho}_{\mathcal{E}_{\text{unital}}})=\text{R}_{\text{n-SB}}(\mathcal{E}_{\text{unital}})$. We arrive at:
\begin{theorem}
The robustness of temporal steerability can be used to quantify unital non-SB channels. 
\end{theorem}

Let us now turn our attention to the temporal Bell scenario (see Fig.~\ref{Fig_temporal_scenarios}), which is equivalent to the LGI scenario. The measurements at both times $t_0$ and $t_1$ are viewed as black boxes with finite inputs $x\in\mathcal{X}$ and $y\in \mathcal{Y}$, respectively. The index $a\in\mathcal{A}$ ($b\in
\mathcal{B}$) is used to denote the measurement outcome at $t_0$ ($t_1$)~\cite{Brukner04,Fritz2010,Emary14}.
The system is assumed to obey the aforementioned macrorealism, which implies that a temporal correlation can be described by a hidden-variable (HV) model, namely
\begin{equation}\label{Eq_temporal_HV}
p(a,b|x,y)\stackrel{\text{MS}}=\sum_{j}p(j)p(a|x,j)p(b|y,j).
\end{equation}
One can see that the hidden parameter $j$ causally determines the well-defined probability distributions $p(a|x, j )$ and $p(b|y, j )$ at times $t_0$ and $t_1$, respectively. 
The violation of the temporal Bell inequality, which has the same form of \eq{Eq_Bell_inequality} but with the correlation obtained from temporally separated measurements, reveals a quantum
correlation effect. This quantum correlation is interpreted as non-macrorealism.
This is because the temporal Bell
inequality is a special kind of LGI, which
is compatible with all macrorealistic physical theories.

In quantum theory, the observed correlation can be described by
\begin{equation}
\begin{aligned}
p(a,b|x,y)&=\text{Tr}\left[( M_{a|x} \otimes M_{b|y})P_{\mathcal{E}}\right]\\
&=\text{Tr}\left[ M_{b|y}\rho_{\mathcal{E}}(a|x)\right].
\end{aligned}
\label{Eq_CJ_pabxy}
\end{equation}
One can note that certifying the
non-macrorealistic probability distribution is mathematically identical to witnessing the locality of the CJ state in \eq{Eq_CJ_pabxy}. 
If the observed correlation admits an HV model, the channel breaks nonlocality for the maximally entangled states by definition. 
Now, we define the robustness of non-macrorealism as 
\begin{equation}\label{Eq_R_MR}
\begin{aligned}
    &\text{R}_{\text{n-MR}}(p(a,b|x,y))=\   \\& \min\left\{ \alpha \Big|
\frac{p(a,b|x,y) + \alpha p'(a,b|x,y)}{1+\alpha}  \in
\mathcal{F}_{\text{HV}} \right\},
\end{aligned}
\end{equation}
where $\mathcal{F}_{\text{HV}}$ denotes the set of temporal correlations admitting an HV model.
We can see that $\text{R}_{\text{n-MR}}$ provides a lower bound for $\text{R}_{\text{n-NLB}}$. Therefore, we have:
\begin{theorem}
The robustness of non-macrorealism can be used to quantify non-NLB channels for the maximally entangled state.
\end{theorem}

Since other works~\cite{Pal2015,Zhang2019} focused on CHSH-NLB channels, we briefly discuss the particular case of the temporal CHSH inequality. Since the unital channel is
CHSH-NLB, when the channel breaks the CHSH nonlocality for the maximally entangled states, the temporal CHSH inequality can be used to certify all unital CHSH-NLB channels. Finally, we emphasize that due to the hierarchy relation of temporal quantum correlations~\cite{Ku2018,Uola2018}, temporal separability implies macrorealism, but not vice versa. Thus, the concept of non-macrorealism can be used to witness non-EB channels. A similar argument for certifying non-SB channels can also be applied for testing non-macrorealism.

\section{Experimental setup and results}\label{Sec_exp}
In this section, we present a proof-of-principle experiment demonstrating (i) the hierarchy of breaking channels and (ii) how temporal quantum correlations can be used to quantify the quantumness of channels. In detail, we consider the 2-dimensional
depolarizing channel, which is a convex combination of white noise
with the input state, namely,
\begin{equation}\label{Eq_depolarizing}
\mathcal{E}_{\text{D}}(v,\tilde{\rho})=v\tilde{\rho}+(1-v)\frac{\openone}{2},
\end{equation}
where $v$ is the mixing parameter. The corresponding PDO can be
expressed as
\begin{equation}
P_{\mathcal{E}_{\text{D}}}=\frac{v}{2}\text{SWAP}+\frac{1-v}{4}\openone,
\end{equation}
In the temporal steering scenario, we consider the three
dichotomic measurements ($\{\hat{X},\hat{Y},\hat{Z}\}$) applied on
the PDO at $t_0$ in order to obtain the maximal temporal
steerability~\cite{Ku20182,ShinLiang2020}. To obtain the robustness of non-macrorealism for the 2-dimensional
depolarizing channel, we consider the two sets
of anti-commuting operators: $\{\hat{X},\hat{Z}\}$ and
$\{(\hat{X}+\hat{Z})/2,(\hat{X}-\hat{Z})/2\}$, which maximizes violation of the temporal CHSH inequality.

\begin{figure}[ht]
\includegraphics[scale=1]{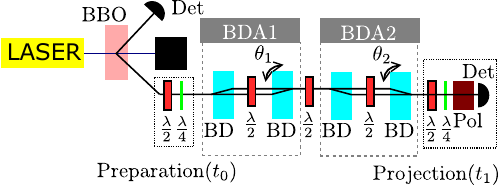}
\caption{\label{fig_setup} Scheme for our experimental
implementation of the depolarizing channel. The
$\beta$-Ba(BO$_2$)$_2$ 
is a nonlinear crystal for spontaneous parametric down-conversion;
$\lambda/2$ and $\lambda/4$ are half- and quarter-wave plates,
respectively; BDs are beam displacers; BDAs are beam-displacer
assemblies; Pol is a polarizer; and Det are single-photon
detectors.}
\end{figure}

We have demonstrated all of these temporal scenarios in a photonic
experiment. The experimental setup is schematically shown in
Fig.~\ref{fig_setup}. In this experiment, qubits are encoded into
the polarization state of individual photons and manipulated using
linear optics. More details on the experimental implementation are
provided in Appendix~\ref{App_experimental_results}.

A quarter and half-wave plates are used to prepare single photons
in the desired polarization state. In our experiment, we prepared
six different initial states which are the eigenstates of operators
$\{\hat{X}, \hat{Y}, \hat{Z}\}$.
This preparation is operationally equivalent to the nondestructive
projective measurement at $t_0$.
The photons then enter the depolarizing channel consisting of two
beam displacer assemblies (BDA), one of which is enveloped by
Hadamard gates ($\hat{H}$). These two BDAs together can perform
one of following operations: $\{\openone, \hat{X}, \hat{Y},
\hat{Z}\}$. We assign each operation a probability depending on the
parameter $v$. To implement the depolarizing channel, we randomly
(with assigned probabilities) with frequency 10 Hz change the
operation and accumulate signal for sufficiently long times (100s)
(see further details in Appendix~\ref{App_experimental_results}).
To analyze the output state, we implement polarization projection
and subsequent detection using a half and quarter-wave plate, a
polarizer and a single photon detector. Note that the
aforementioned half-wave plate is used to implement both the
second Hadamard gate and the analysis. {All combinations of input states together with projections onto the eigenstates of the $\{\hat{X}, \hat{Y}, \hat{Z}\}$ operators allows to perform a full process tomography which characterizes the entire channel in terms of the PDO.} 

\begin{figure}[tbp]
\includegraphics[width=1\columnwidth]{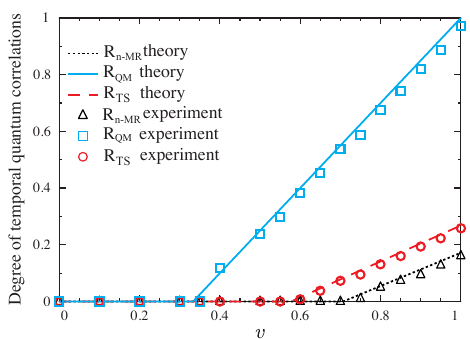}
\caption{Effects of the 2-dimensional depolarizing channel,
characterized by the mixing parameter $v$ in \eq{Eq_depolarizing},
on the robustness-based measures including: quantum memory (blue squares and solid line), temporal steerability (red circles and dashed line) and non-macrorealism (black triangles and dotted line). The experimental results are marked
by symbols. The vanishing parameter
of the robustness of quantum memory is $1/3$, which distinguishes the EB channel in the quantum domain. Moreover, for the temporal steering and CHSH
scenarios, the vanishing parameters are respectively $1/\sqrt{3}$
and $1/\sqrt{2}$, which are the same as the boundaries of the
2-dimensional depolarizing channels for the SB and
CHSH-NLB channels under the three and two dichotomic
measurement settings.}
\label{fig_experimental_results}\centering %
\end{figure}

Our experimental results are plotted in
Fig.~\ref{fig_experimental_results}. As can be seen, the PDO can be used to measure quantum memory in the form of the robustness-based measure. Moreover, the PDO is separable when $v\leq 1/3$, which saturates the bound of the EB channel in the quantum domain~\cite{Rosset2018}. For the temporal
steering and temporal Bell scenarios, each value in Fig.~\ref{fig_experimental_results} provides a lower bound on the robustness of the non-SB and the non-NLB channels. One can see that for the depolarizing channel,
the vanishing parameter $v$ of the
SB channel in the quantum domain is
$v=1/\sqrt{3}$. This vanishing parameter under
the three measurements setting scenario is identical to that in the
$3$-incompatibility-breaking channel, which breaks the
incompatibility of every collection of three
measurements~\cite{Heinosaari2015}. In other words, we can also
say that it is the $3$-SB channel which breaks the spatial
steerability under all of the three measurement settings.
Finally, the robustness of the non-NLB channel suggests that under the two binary inputs scenario, the vanishing parameter is $v=1/\sqrt{2}$, which is
identical to the boundary of the CHSH-NLB channel under the
2-dimensional depolarizing channel~\cite{Pal2015,Zhang2019}. Our experimental results also show the hierarchy between breaking channels (see also Fig.~\ref{fig:breaking}). The error of all experimentally obtained quantities is estimated by assuming
the Poisson distribution of the photon counts. Errors of
quantities obtained from the density matrices is determined by a
Monte-Carlo method. Further details are given in Appendix~\ref{App_experimental_results}.

\section{Discussion and conclusions}\label{sec:Discussion}

In this work, we have proposed the steerability-breaking (SB)
channel, which is defined in an analogous way to the
entanglement-breaking (EB) and nonlocality-breaking (NLB)
channels. We have then proven a strict hierarchy between these concepts and experimentally illustrated it with the qubit depolarizing channel in a photonics system.
We then proposed the robustness-based measures to quantify the degree of different types of non-breaking channels.
In the Heisenberg picture, we can formally show that many well-known aspects concerning the incompatibility-breaking channels are equivalent with the corresponding ideas in the SB channel, including the mathematical description and robustness-based measures.

We have also connected the non-breaking channels with non-macrorealism, which is the basis of the Leggett-Garg inequality, normally used to check for quantum effects in macroscopic systems.
More specifically, we have shown that the robustness-based measures of the non-EB and non-SB channels can be accessed by temporal nonseparability and channel steerability, respectively.
The temporal steerability and non-macrorealism can be applied to
quantify the unital non-SB channel and the non-NLB
channel for the maximally entangled state. We also showed that all unital non-CHSH-NLB channels which only break the CHSH nonlocality instead of the
general nonlocality can be certified by the temporal CHSH inequality. Therefore, the above breaking channels can be
quantified in the temporal domain without an entangled source. We have also demonstrated the photonics experiment to
explicitly show how temporal quantum correlations can be used to
quantify the non-breaking channels.

Several natural questions can be discussed: Can all non-NLB
channels be certified in the temporal scenario?
Similar to the temporal semi-quantum game, which certifies all
non-EB channels in the measurement-device-independent scenario, the
measurement-device-independent channel steering has been
proposed~\cite{Jeon2020}, but without considering the relationship with
SB channels. Can measurement-device-independent channel steering certify all non-SB channels? \red{A recently proposed loophole-free test of macrorealism~\cite{Joarder2022PRXQ} motivates us to ask whether we can generalize their setup and extend to a loophole-test of non-breaking channels with temporal quantum correlations?} Is the negativity-like measure using the temporal nonseparability a useful quantum-memory monotone? It has been shown that the negativity-like measure of the temporal nonseparability can be used to quantify quantum causality~\cite{Fitzsimons2015} and to estimate channel capacity~\cite{Pisarczyk2019}. We have partially addressed this question by \red{showing that the negativity-like measure of the temporal nonseparability is a lower bound of the negativity of a quantum channel~\cite{Pisarczyk2019}, for which we prove that it satisfies the conditions of a quantum memory monotone under the assumption in Appendix~\ref{App_f_function}. In other words, the negativity-like measure of the temporal nonseparability can be used to quantify quantum memory.}

\begin{acknowledgments}
The reported experiment was performed by J.K. under the guidance of A.\v{C}. and K.L.
The authors acknowledge Karol Bartkiewicz for his help in
numerical postprocessing of our experimental data. This work is
supported partially by the National Center for Theoretical
Sciences and Ministry of Science and Technology, Taiwan, Grants
No. MOST 110-2123-M-006-001, and MOST
109-2627-E-006-004, and the Army Research Office (under Grant No.
W911NF-19-1-0081). H.Y. acknowledges partial support from the National Center for Theoretical
Sciences and Ministry of Science and Technology, Taiwan, Grants No. MOST 110-2811-M-006-546.
MTQ acknowledges the Austrian Science Fund (FWF) through the SFB project BeyondC (sub-project F7103), a grant from the Foundational Questions Institute (FQXi) as part of the Quantum Information Structure of Spacetime (QISS) Project (qiss.fr). This project has received funding from the European Union’s Horizon 2020 research and innovation programme under the Marie Sk\l odowska-Curie grant agreement No, 801110 and the Austrian Federal Ministry of Education, Science and Research (BMBWF). It reflects only the authors' view, the EU Agency is not responsible for any use that may be made of the information it contains.
 N.L. acknowledges partial support from JST
PRESTO through Grant No. JPMJPR18GC. A.M. is supported by the
Polish National Science Centre (NCN) under the Maestro Grant No.
DEC-2019/34/A/ST2/00081.
S.-L. C. acknowledges support from the Ministry of Science and Technology Taiwan (Grant No. MOST 110-2811-M-006-539).
F.N. is supported in part by: Nippon
Telegraph and Telephone Corporation (NTT) Research, the Japan
Science and Technology Agency (JST) [via the Quantum Leap Flagship
Program (Q-LEAP) program, the Moonshot R\&D Grant Number
JPMJMS2061, the Japan Society for
the Promotion of Science (JSPS) [via the Grants-in-Aid for
Scientific Research (KAKENHI) Grant No. JP20H00134], the Army Research Office
(ARO) (Grant No. W911NF-18-1-0358), the Asian Office of Aerospace
Research and Development (AOARD) (via Grant No. FA2386-20-1-4069),
and the Foundational Questions Institute Fund (FQXi) via Grant No.
FQXi-IAF19-06. J.K. and K.L. acknowledge funding by the Czech Science
Foundation (Grant No. 20-17765S) and by Palack\'y University
(IGA\_PrF\_2021\_004).
\end{acknowledgments}

\appendix
\section{Robustness-based measures of breaking channels} \label{App_monotones}
We first briefly introduce the concept of a resource theory. In general, a resource theory is consisted of three components: (i) a set of free states $\mathcal{F}$, (ii) a set of free operations $\mathcal{O}$, and (iii) resource measures $Q$. Here, we only consider the most general free operations (resource-non-generating maps) and denote them as the resource-free operations for convenience. They map free states to free states. Once the free states and operations are established, we can define a resource measure (or a resource monotone)  $Q$ , which obeys:
\begin{itemize}
\item[(i)] It vanishes for any free state:
\begin{equation}
Q(\mathcal{E})=0\quad \text{if}\quad   \mathcal{E}\in\mathcal{F} .
\end{equation}
\item[(ii)] It is non-increasing under free operations, i.e.,
\begin{equation}
 Q\left(\Lambda(\mathcal{E}) \right)\leq Q(\mathcal{E}),
\end{equation}
where $\Lambda$ is a free operation.
\item [(iii)] Given a distribution $p(i)$ and resources $\mathcal{E}_i$, we say that $Q$ is a convex monotone if it satisfies the inequality
\begin{equation}
\begin{aligned}
Q\left(\sum_ip(i)\mathcal{E}_i\right)
\leq \sum_i p(i) Q(\mathcal{E}_i).
\end{aligned}
\end{equation}
\end{itemize}
\red{In a resource theory, the robustness-based measure is often used because the quantity itself also directly tells us the degree of advantages that the underlying quantum object provides  (see how it can be applied to general resource theories~\cite{Chitambar2019RMP,Ryuji2019,Tan20212} and some explicit applications to quantum theories ~\cite{Ryuji2020,Buscemi2020,Skrzypczyk2019,Ku2022}). By definition, a robustness-based measure denotes how much noise can be added to the input resource before the output resource becomes free.}
In Table.~\ref{tab:notations}, we summarize the notations of the robustness-based measure, \red{that we considered}.
In what follows, we apply the above concepts to entanglement-breaking (EB), nonlocality-breaking (NLB), and steerability-breaking (SB) channels.

\begin{table}[h]
\caption{\label{tab:notations} Table of symbols of different robustness-based measures. The first column represents different quantum objects/phenomena discussed in our work. The second column describes the symbols used for representing the associated robustness-based measures. The third column indicates the equations where the measures appear.}

\begin{ruledtabular}

\begin{tabular}{lccc}
quantum objects & symbols of measures & appearance\\
\hline
quantum memory                            & $\text{R}_{\text{ QM}}$ & \eq{Eq_R_QM}\\
non-nonlocality breaking channel  & $\text{R}_{\text {n-NLB}}$ & \eq{Eq_R_NLB}  \\
non-steerability breaking channel & $\text{R}_{\text {n-SB}}$  & \eq{Eq:14}\\
channel steering                               & $\text{R}_{\text {CS}}$      &\eq{Eq_CSR} \\
temporal steering                      & $\text{R}_{\text {TS}}$      & \eq{Eq_TSR}\\
non-macrorealism                            & $\text{R}_{\text {n-MR}}$ & \eq{Eq_R_MR}\\

\end{tabular}
\end{ruledtabular}
\end{table}

\subsection{Quantum memory}

As mentioned in the main text, the free states of a resource theory of quantum memory is the set of EB channels~\cite{Rosset2018,Xiao2021}. In Ref.~\cite{Xiao2021}, it has been shown that the robustness of the quantum memory $\text{R}_{\text{QM}}$ defined in \eq{Eq_R_QM} satisfies the quantum memory monotone, in the sense that it is a non-increasing function under quantum-memory free operations. Readers can find a more detailed discussion in Ref.~\cite{Xiao2021}. Instead of the most general free operations, we discuss a particular choice of free operations proposed in Appendix~\ref{App_f_function}.

\subsection{Nonlocality-breaking channel}

Here, we first define the most general free operations within the resource theory of non-NLB-breaking channels, as the non-NLB free operations $\Lambda_{\text{NLB}}$. It is the physical transformation of quantum channels (quantum supermap) that maps NLB channels into (same or other) NLB channels only,
\begin{equation}\label{Eq_freeO_NLB}
    \mathcal{O}_{\text{NLB}}=\{\Lambda_{\text{NLB}}:\Lambda_{\text{NLB}}(\mathcal{E})\in \mathcal{F}_{\text{NLB}},\forall \mathcal{E}\in\mathcal{F}_{\text{NLB}}\}.
\end{equation}
By defining a noisy channel, $E_{\alpha}=\frac{\mathcal{E} + \alpha \mathcal{E'}}{1+\alpha}$, with $\mathcal{E}'$ being any channel, we can recall that the robustness of a non-NLB channel is defined as
\begin{equation}\label{Eq_R_NLB_app}
\begin{aligned}
&\text{R}_{\text{n-NLB}}(\mathcal{E}) = \min\{\alpha\geq 0\Big|\\& \text{Tr}\left[(M_{a|x}\otimes M_{b|y})(E_{\alpha}\otimes \openone)\tilde{\rho}_{\text{A,B}}\right]\in \mathcal{F}_{\text{LHV}}~\forall M_{a|x},M_{b|y},\tilde{\rho}_{\text{A,B}} \}.
\end{aligned}
\end{equation}
Here, $\mathcal{F}_{\text{LHV}}$ is a set of local correlations admitting the HV model defined in \eqref{Eq_HV}.
By definition, one can see that the optimal channel $\mathcal{E}_{\alpha^*}$ is a NLB-breaking channel.
In what follows, we show that the robustness of a non-NLB channel, defined in \eqref{Eq_R_NLB}, is a non-NLB monotone.
\begin{lemma}
The value of the robustness of a non-NLB channel is zero when the channel is NLB.
\end{lemma}
\emph{Proof.---} This follows directly from the definition in \eq{Eq_R_NLB_app}. $\square$

\begin{lemma}
The robustness of a non-NLB channel is decreasing under the non-NLB free operations.
\end{lemma}
\emph{Proof.---} We first denote the optimal solution in \eqref{Eq_R_NLB_app} with an asterisk. Since the non-NLB free operation is linear, after applying it to $\mathcal{E}^*_{\alpha}$, we have
\begin{equation}
    \Lambda\left(\frac{\mathcal{E} + \alpha^* \mathcal{E'}}{1+\alpha^*}\right)=\frac{\Lambda(\mathcal{E})}{1+\alpha^*}+\frac{\alpha^*\Lambda(\mathcal{E'})}{1+\alpha^*}=\Lambda(\mathcal{E}_{\alpha^*})\in \mathcal{F}_{\text{NLB}}.
\end{equation}
Here, we use the property of the non-NLB operation in \eqref{Eq_freeO_NLB}.
Because $\Lambda(\mathcal{E}_{\alpha^*})$ is still an NLB channel, the constraints in \eqref{Eq_R_NLB_app} are always satisfied. However, $\alpha^*$ is not necessarily the minimum of $\text{R}_{\text{n-NLB}}(\Lambda(\mathcal{E}))$. Therefore, we conclude that $\text{R}_{\text{n-NLB}}(\Lambda(\mathcal{E}))\leq \text{R}_{\text{n-NLB}}(\mathcal{E})$. $\square$

\begin{lemma}
The robustness of the non-NLB channel satisfies convexity, namely
\begin{equation}
    \text{R}_{\text{n-NLB}}\left(\sum_i p(i)\mathcal{E}_i\right)\leq \sum_i p(i)\text{R}_{\text{n-NLB}}(\mathcal{E}_i).
\end{equation}
\end{lemma}
\emph{Proof.---} For each $i$, the solution in \eqref{Eq_R_NLB_app} is denoted by $\alpha_i^*$ and the channel is denoted as $\mathcal{E}'_i$. According to the definition in Eq. \eqref{Eq_R_NLB_app}, we have
\begin{equation}\label{Eq_convex_NLB_1}
\frac{\mathcal{E}_i}{1+\alpha_i^*} + \frac{\alpha_i^* \mathcal{E}'_i}{1+\alpha_i^*}=\mathcal{E}_{\alpha_i^*}\in\mathcal{F}_{\text{NLB}}.
\end{equation}
We now define the coefficients $\alpha=\sum_i p(i)\alpha_i^*$, $\mathcal{E}=\sum_i p(i)\mathcal{E}_i$, and $\mathcal{E}'=1/\alpha\sum_i p(i)\alpha_i^*\mathcal{E}'_i\in\mathcal{F}_{\text{NLB}}$, and we obtain
\begin{equation}\label{Eq_convex_NLB_2}
    \begin{aligned}
    \frac{1}{1+\alpha}\mathcal{E}+\frac{\alpha}{1+\alpha}&\mathcal{E}'=\frac{1}{1+\alpha}
    \left(\sum_i p(i)\mathcal{E}_i+\sum_i p(i)\alpha_i^*\mathcal{E}'_i  \right)\\
    &=\sum_i p(i)\frac{1}{1+\alpha}
    \left(\mathcal{E}_i+ \alpha_i^*\mathcal{E}'_i  \right)\\
    &=\sum_i p(i)\frac{1}{1+\alpha}
    \left(\alpha_i^*+1  \right)\mathcal{E}_{\alpha_i^*}\in\mathcal{F}_{\text{NLB}}.
    \end{aligned}
\end{equation}
The last equality holds due to \eqref{Eq_convex_NLB_1} and the classical mixture of NLB channels is still an NLB channel.
Finally, we complete the proof as follows:
\begin{equation}
    \text{R}_{\text{n-NLB}}\left(\sum_ip(i)\mathcal{E}_i\right)=\text{R}_{\text{n-NLB}}(\mathcal{E})\leq \alpha=\sum_i p(i)\text{R}_{\text{n-NLB}}(\mathcal{E}_i),
\end{equation}
The inequality holds due to the definition in \eq{Eq_R_NLB_app} and \eq{Eq_convex_NLB_2}.$\square$

\subsection{Steerability-breaking channel}

Similar to the NLB scenario, we can define the most general free operations within the resource theory of non-SB-breaking channels as non-SB free operations in the sense that any of them maps an SB channel into another SB channel, namely
\begin{equation}\label{Eq_freeO_SB}
    \mathcal{O}_{\text{SB}}=\{\Lambda_{\text{SB}}:\Lambda_{\text{SB}}(\mathcal{E})\in \mathcal{F}{\text{SB}},\forall \mathcal{E}\in\mathcal{F}_{\text{SB}}\}.
\end{equation}
Since we have shown that the SB channels are the same as incompatible-breaking channels in the Heisenberg picture, the dual maps of the non-SB free operation are naturally non-incompatible-breaking free operations (see also Appendix~\ref{App_SB_IB_are_the_same}).
We then reference the robustness of a non-SB channel in the 
Heisenberg picture given by \eq{Eq_nSBR}.
We now can show that the robustness of the non-SB channel defined in \eq{Eq_nSBR} satisfies the requirements of a non-SB monotone.
\begin{lemma}
The value of the robustness of a non-SB channel is zero when the channel is SB.
\end{lemma}
\emph{Proof.---} This follows directly from the definition. $\square$

\begin{lemma}
The robustness of the non-SB channel is decreasing under the non-SB free operations.
\end{lemma}
\emph{Proof.---} The proof is similar to that of the monotonicity of the robustness of the non-NLB channel. Since we have shown that SB and incompatible-breaking channels are equivalent, in what follows, we prove the monotonicity in the Heisenberg picture.
With the linearity of the non-SB free operation, we reformulate the constraints in \eq{Eq_nSBR} as
\begin{equation}
\frac{\Lambda^\dagger(\mathcal{E}^\dagger) + \alpha \Lambda^\dagger(\mathcal{E'}^\dagger)}{1+\alpha}=\Lambda^\dagger(\mathcal{E}^\dagger_{\text{JM}}) \in \mathcal{F}_{\text{JM}}.
\end{equation}
Here, we use the property that of the dual non-SB free operation is a non-incompatible free operation.
Because $\Lambda^\dagger(\mathcal{E}_{\text{JM}}^\dagger)$ is still an incompatible-breaking channel, the constraints in \eq{Eq_nSBR} are always satisfied, while it may not be the minimizer for $\text{R}_{\text{n-SB}}(\Lambda(\mathcal{E}))$. Thus, we can conclude that $\text{R}_{\text{n-SB}}(\Lambda(\mathcal{E}))\leq \text{R}_{\text{n-SB}}(\mathcal{E})$. $\square$

\begin{lemma}
The robustness of the non-SB channel satisfies convexity, namely
\begin{equation}
    \text{R}_{\text{n-SB}}\left(\sum_i p(i)\mathcal{E}_i\right)\leq \sum_i p(i)\text{R}_{\text{n-SB}}(\mathcal{E}_i)
\end{equation}
\end{lemma}
\emph{Proof.---} The proof is analogous to that of the convexity of the robustness of the non-NLB channel, therefore we do not repeat it here.$\square$

\section{The set of the steerability-breaking channels is identical to the set of the incompatibility-breaking channels.}\label{App_SB_IB_are_the_same}

We first summarize that the following statements are equivalent
related to a quantum channel $\mathcal{E}$:

\begin{itemize}
\item[\#1.] A quantum channel breaks the steerability for an arbitrary quantum state $\tilde{\rho}_{A,B}$ and an arbitrary measurement set $\{M_{a|x}\}$.
\item[\#2.] A quantum channel breaks the steerability for the maximally entangled states and an arbitrary measurement set $\{M_{a|x}\}$.
\item[\#3.] A dual of the quantum channel $\mathcal{E}^{\dagger}$ breaks the incompatibility for the arbitrary measurement set $\{M_{a|x}\}$. In other words, $\mathcal{E}^{\dagger}(M_{a|x})$ is jointly measurable.
\end{itemize}

\begin{lemma}
The above-mentioned statements \#2 and \#3 are equivalent.
\end{lemma}
\emph{Proof.---}For \#3 $\Rightarrow$ \#2, it is trivial because the
incompatible measurement is a necessary condition for
demonstrating the steerability~\cite{Uola2014,Quintino14}. If the
channel is incompatibility-breaking, it must also be steerability
breaking. For \#2\;$\Rightarrow$\;\#3, we consider the state
$\tilde{\rho}_{A,B}=\ket{\Phi}\bra{\Phi}$, with
$\ket{\Phi}=\frac{1}{d}\sum_{i=1}^d\ket{i}\otimes\ket{i}$, and an
arbitrary measurement set $\{M_{a|x}\}$. The corresponding
assemblage can be obtained by
\begin{equation}\label{Eq_B1}
\begin{aligned}
\text{Tr}_{A}&\left[(M_{a|x}\otimes \openone) (\mathcal{E}\otimes\openone)\ket{\Phi}\bra{\Phi}  \right]\\&=\text{Tr}_{A}\left[(\mathcal{E}^{\dagger}(M_{a|x})\otimes \openone) \ket{\Phi}\bra{\Phi}  \right]\\
&=\frac{1}{d}\left[\mathcal{E}^{\dagger}(M_{a|x})\right]^{\text{T}}.
\end{aligned}
\end{equation}
Since the assemblage itself must satisfy an LHS model, we can
reformulate the above as
\begin{equation}\label{Eq_from_2_to_3}
\begin{aligned}
\left[\mathcal{E}^{\dagger}(M_{a|x})\right]^{\text{T}}&=d\times\sum_{\lambda}p(a|x,\lambda)p(\lambda)\tilde{\rho}_{\lambda}\\
&=\sum_{\lambda}p(a|x,\lambda)M_{\lambda}.
\end{aligned}
\end{equation}
By summing up the outcome $a$ in \eq{Eq_from_2_to_3},
\begin{equation}
\begin{aligned}
\openone\equiv\sum_{a}\left[\mathcal{E}^{\dagger}(M_{a|x})\right]^{\text{T}}&=\sum_{a,\lambda} p(a|x,\lambda)M_{\lambda}\\
&=\sum_{\lambda} M_{\lambda},
\end{aligned}
\end{equation}
we can show that $M_{\lambda}=d p(\lambda)\tilde{\rho}_{\lambda}$ is a
valid POVM. Here, we use the facts that
$\left[\mathcal{E}^{\dagger}(M_{a|x})\right]^{\text{T}}$ is a
valid POVM and
$\sum_{a}\left[\mathcal{E}^{\dagger}(M_{a|x})\right]^{\text{T}}=\openone$.
Therefore, the last equation in \eq{Eq_from_2_to_3} is jointly
measurable. From the above, we prove \#2\;$\Leftrightarrow$\;\#3. $\square$

\begin{lemma}
The above-mentioned statements \#1 and \#3 are equivalent.
\end{lemma}
\emph{Proof.---} Since \#3\;$\Rightarrow$\;\#1 is trivial, we are going
to show \#1$\Rightarrow$ \#3. An equivalent description of
\#1\;$\Rightarrow$\;\#3 is $(\neq$ \#3)\;$\Rightarrow$ $(\neq$ \#1).
According to the definition of the incompatibility-breaking
channel, there exists a set of measurements $\{M_{a|x}\}$ such
that the ``evolved" measurement
$\{\mathcal{E}^{\dagger}(M_{a|x})\}$ is incompatible. Now, if we
consider Alice and Bob sharing a maximally entangled state
$\tilde{\rho}_{A,B}=\ket{\Phi}\bra{\Phi}$, we can
obtain the following assemblage, which is the same as \eq{Eq_B1}.
Since
$\frac{1}{d}\left[\mathcal{E}^{\dagger}(M_{a|x})\right]^{\text{T}}$
must be steerable, the channel is not a SB channel by definition.
$\square$

Although from the above two Lemmas, it is enough to show that the
statements \#1, \#2, and \#3 are equivalent. We still provide an
independent proof of \#2\;$\Leftrightarrow$\;\#3.
\begin{lemma}
The above-mentioned statements \#1 and \#2 are equivalent.
\end{lemma}
\emph{Proof.---} By definition, \#1\;$\Rightarrow$\;\#2 is trivial. In
the following, we show that \#2\;$\Rightarrow$\;\#1 also holds. First, we
introduce the maximally entangled state
$\ket{\Phi}\bra{\Phi}=\frac{1}{d}\sum_{i,j}\ket{i}\bra{j}\otimes\ket{i}\bra{j}$,
the hidden state
$\tilde{\rho}^\lambda=\sum_{i,j}\chi^{\lambda}_{i,j}\ket{i}\bra{j}$,
and the arbitrary bipartite quantum state
$\tilde{\tau}=\sum_{mnpq}\Upsilon^{m,n}_{p,q}\ket{m}\bra{n}\otimes\ket{p}\bra{q}$
in the matrix representation. Here $\chi^\lambda_{i,j}$ and
$\Upsilon_{mnpq}$ are the entries of the corresponding states.
From \#2, there must exist a hidden state model for the channel
$\mathcal{E}$ breaking the steerability of the maximally entangled
state for an arbitrary measurement set $\{E_{a|x}\}$ and we have
\begin{equation}
\begin{aligned}
&\text{Tr}_{A}\left[(E_{a|x}\otimes\openone)(\mathcal{E}\otimes\openone)\ket{\Phi}\bra{\Phi}\right]\\&=\frac{1}{d}\sum_{i,j}\text{Tr}\left[E_{a|x}\mathcal{E}(\ket{i}\bra{j})\right]\ket{i}\bra{j}\\
&=\sum_{\lambda}p(a|x,\lambda)p(\lambda)\tilde{\rho}^\lambda\\
&=\sum_{i,j,\lambda}p(a|x,\lambda)p(\lambda)\chi_{i,j}^\lambda\ket{i}\bra{j}.
\end{aligned}
\end{equation}
By linearity, we obtain
\begin{equation}
\text{Tr}\left[E_{a|x}\mathcal{E}(\ket{i}\bra{j})\right]=d\sum_{\lambda}p(a|x,\lambda)p(\lambda)\chi^\lambda_{i,j}.
\end{equation}
Now, inserting the arbitrary bipartite state into the definition
of the SB channel, one finds
\begin{equation}
\begin{aligned}
\text{Tr}_{A}&\left[\mathcal{E}^{\dagger}(E_{a|x})\otimes\openone\tilde{\tau}\right]=\sum_{m,n,p,q}\Upsilon^{m,n}_{p,q}\text{Tr}\left[E_{a|x}\mathcal{E}(\ket{m}\bra{n})\right]\ket{p}\bra{q}\\
&=\sum_{m,n,p,q}\Upsilon^{m,n}_{p,q}\left(d\sum_{\lambda}p(a|x,\lambda)p(\lambda)\chi^\lambda_{i,j}\right)\ket{p}\bra{q}\\
&=d\sum_{\lambda}p(a|x,\lambda)p(\lambda)\sum_{m,n,p,q}\Upsilon^{m,n}_{p,q}\chi^\lambda_{i,j}\ket{p}\bra{q}\\
&=d\sum_{\lambda}p(a|x,\lambda)p(\lambda)\text{Tr}_{A}\left[\tilde{\tau}(\openone\otimes
(\tilde{\rho}^\lambda)^{\text{T}})\right].
\end{aligned}
\label{Eq_2_to_1}
\end{equation}
It is trivial to see that
$\text{Tr}_{A}\left[\tilde{\tau}(\openone\otimes
(\tilde{\rho}^\lambda)^{\text{T}})\right]$ is positive semidefinite. Now
the remaining problem is to show that
$d\text{Tr}_{A}\left[\tilde{\tau}(\openone\otimes
(\tilde{\rho}^\lambda)^{\text{T}})\right]$ is a valid quantum state. Since
the LHS of \eq{Eq_2_to_1} is a valid assemblage, the marginal
assemblage is a valid quantum state with unit trace. Thus, we
trace the marginal assemblage on the RHS and obtain
\begin{equation}
\begin{aligned}
&\sum_a\text{Tr}\left[(E_{a|x}\otimes\openone)(\mathcal{E}\otimes\openone)\tilde{\tau}\right]\\&=d\sum_{a,\lambda}p(a|x,\lambda)p(\lambda)\text{Tr}\left[\tau(\openone\otimes (\tilde{\rho}^\lambda)^{\text{T}})\right]\\
&=d\sum_{\lambda}p(\lambda)\text{Tr}\left[\tilde{\tau}(\openone\otimes
(\tilde{\rho}^\lambda)^{\text{T}})\right]\equiv 1.
\end{aligned}
\end{equation}
Since $\sum_{\lambda}p(\lambda)=1$, and
$\text{Tr}\left[\tilde{\tau}\right]=\text{Tr}\left[(\tilde{\rho}^\lambda)^{\text{T}}\right]=1$,
the only choice for satisfying the above relations is
$d\text{Tr}\left[\tau(\openone\otimes
(\tilde{\rho}^\lambda)^{\text{T}})\right]=1$, with arbitrary states
$\tilde{\tau}$ and $(\tilde{\rho}^\lambda)^{\text{T}}$. Thus,
$d\text{Tr}_{A}\left[\tilde{\tau}(\openone\otimes
(\tilde{\rho}^\lambda)^{\text{T}})\right]$ is a valid assemblage. With the
above, we complete \#2\;$\Leftrightarrow$\;\#3.$\square$

\section{From PDO to the temporal semi-quantum game}\label{App_temporal_semi}
The temporal semi-quantum game considers two players, Alice and
Bob. They are able to generate the same set of characterized
quantum states $\{\tilde{\sigma}_x\}\in \mathcal{D}(H_{t_0}) $ and
$\mathcal{D}(H_{t'_1})$ with the finite number set $x\in\mathcal{X}$
and $y\in\mathcal{Y}$. Alice first sends a state
$\tilde{\sigma}_x$ from the set into a quantum channel
$\mathcal{E}$. After the channel, Bob performs a joint measurement
$B_b$ with the evolved state and the second quantum input sending
from Bob. Here, $b$ is the observed measurement outcome. To
certify all non-EB channels, the set of quantum states should form
a tomographically complete set, e.g., the eigenstates of the Pauli
matrices.

To show the results in this section, it is convenient to
compactly reformulate the joint measurement with the quantum
input at $t_1$ as
$M_{b}=\text{Tr}_{t'_1}[B_b(\openone\otimes\tilde{\sigma}_y)]$, which
forms an effective POVMs. Following Born's rule, the
probability distribution obtained can be expressed as
\begin{equation}
p(b|x,y)=\text{Tr}\left[B_b\tilde{\sigma}_y\otimes\mathcal{E}(\tilde{\sigma}_x)\right]=\text{Tr}\left[M_{b}\mathcal{E}(\tilde{\sigma}_x)\right].
\end{equation}

Now, we show that the temporal semi-quantum game is operationally equivalent with
the causality quantity PDO $P_{\mathcal{E}}$. First, we recall
that the normalized quantum states at time $t_1$ can be obtained
by applying a set of measurements $\{M_{x}\}$ at $t_0$ in
\eq{Eq_from_PDO2assemblage} with normalization~\cite{Ku2018}. More
specifically, the effect of the measurement at $t_0$ on PDO
generates the evolved postmeasurement states at $t_1$. Finally, the
measurement $M_{b}$ at $t_1$ is implemented, and we can obtain the
same probability distribution in the temporal semi-quantum game.

\section{Proof of the hierarchy in quantum non-breaking channels}~\label{App_hierarchy}

According to the definitions in Eqs.~\eqref{Eq_NLB_channel} and~\eqref{Eq_SB_channel}, one can trivially see that the set of all
SB channels must be NLB~\cite{Wiseman2007}. Thus,
we have $\mathcal{F}_{\text{SB}}\subseteq\mathcal{F}_{\text{NLB}}$. To show that the hierarchy is indeed strict, we consider the
single-qubit depolarizing channel defined in Eq.~\eqref{Eq_depolarizing}. It is known that the qubit depolarizing channel is EB if and only if $v\leq1/3$ and that it is CHSH-NLB if and only if $v\leq 1/\sqrt{2}$ \cite{Pal2015}.
Now, we introduce the two-qubit isotropic state 
\begin{equation}
\rho_{\text{ISO}}(v)=v\ket{\phi^+}\bra{\phi^+} + (1-v) \openone/4,
\end{equation} 
which is steerable (with projective measurements) when $v>1/2$~\cite{Wiseman2007}; hence,  due to Theorem~\ref{theory_SB_maximally_input}, the qubit isotropic channel is not SB for $v>1/2$. Also, for $v\leq 5/12$, the isotropic state is unsteerable for any POVM~\cite{Quintino2015,Heinosaari2015}; hence, $v\leq5/12$ ensures SB.

The two-qubit isotropic state violates Bell's inequality when $v>0.696$~\cite{Divinszky2017}. Hence the depolarizing channel cannot be NLB in this range. In order to prove that a quantum channel is NLB, we use a local hidden variable for quantum measurements~\cite{Quintino2016}. Reference \cite{Hirsch2018} shows that for any set of qubit measurements $\{M_{a|x}\}$, when  $v\leq 0.525,$ the set of measurements,
\begin{equation}
\{v M_{a|x} +(1-v)\text{Tr}(M_{a|x})\openone/2\},
\end{equation}
admits a local hidden variable model. That is, for any bipartite quantum state, when $v\leq 0.525$, if Alice performs this set of noise measurements, the statistics are necessarily Bell local. Since the depolarizing channel is self-dual $(\mathcal{E}_{D}=\mathcal{E}_{D}^\dagger)$, the qubit depolarizing channel is NLB when $v\leq 0.525$.
$\square$

\section{Experimental implementation details and data}\label{App_experimental_results}

As a source of horizontally polarized discrete photons, we use a
type-I process of spontaneous parametric down-conversion. One
photon from each pair serves as a herald while the other one
enters the experimental setup. More specifically, a Coherent
Paladin laser at 355~nm is used to pump a $\beta$-Ba(BO$_2$)$_2$
crystal which produces two photons at 710~nm. In this experiment,
we employ polarization encoding associating the horizontal and
vertical polarization with the logical states $\ket{0}$ and
$\ket{1}$, respectively.

For realizing the depolarizing channel, we consider the following
Kraus representation,
\begin{equation}\label{Eq_depolarizing_experimental}
\tilde{\mathcal{E}}_{\text{D}}(p,\tilde{\rho})=
p\tilde{\rho}+\frac{(1-p)}{3} (\hat{X}\tilde{\rho}\hat{X} +
\hat{Y}\tilde{\rho}\hat{Y} + \hat{Z}\tilde{\rho}\hat{Z}),
\end{equation}
where $p$ parameterizes the channel. As one can see, there is a
linear transformation between the above formula and the one in
Eq.~\eqref{Eq_depolarizing} via $p=(3v+1)/4$. The depolarizing
channel (Fig.~\ref{fig_setup}) is implemented by means of two beam
displacer assemblies (BDA1 and BDA2). A beam displacer assembly
consists of two beam displacers and a half-wave plate in between.
The horizontally and vertically polarized parts of the wave packet
are displaced on the first beam displacer, then the polarizations
are swapped on the half-wave plate and finally beams are rejoined
at the second beam displacer. By slightly tilting the second beam
displacer, which is mounted on a piezo actuator, we introduce a
phase shift $\theta_i$ (where $i$ indexes the two beam displacer
assemblies) between the horizontal and vertical component of the
photon's polarization state.

The role of the first beam displacer assembly (BDA1) is to
implement either the $\openone$ or the $\hat{Z}$ (phase flip)
operation depending on whether the introduced phase shift is
$\theta_1 = 0$ or $\theta_1=\pi$, respectively. The second beam
displacer assembly (BDA2) is enveloped by two half-wave plates
(rotated by 22.5° with respect to the horizontal polarization),
each serving as a Hadamard gate ($\hat{H}$). The overall effect of
the second beam displacer together with these gates is the
implementation of either $\openone$ or the $\hat{X}$
$=\hat{H}\hat{Z}\hat{H}$ (bit flip) operation,  provided we set
the phase shift to $\theta_2=0$ or $\theta_2 = \pi$, respectively.
Note that the remaining operation is implemented as a product of
both beam displacer assembly actions $\hat{Y}=i\hat{X} \hat{Z}$
($\theta_1=\theta_2=\pi$).

To depolarize the photon state as prescribed in
Eq.~\eqref{Eq_depolarizing_experimental} we generate, with
frequency $f_r$, a random real number $r \in [0,1]$ (uniformly
distributed). Based on the value of $r$, we choose the setting of
$\theta_1$ and $\theta_2$ (see Table~\ref{table_p_parameter}).

To analyze the output state, we implement the polarization
projection and subsequent detection using a half and quarter-wave
plate, a polarizer, and a single-photon detector. Note that the
aforementioned half-wave plate is used to implement both the
second Hadamard gate and the analysis. The measured signal, for
any combination of prepared and projected states, is integrated
over a period $T\gg 1/f_r$. In our experiment, we accumulate the
number of heralded photon detections for $T=100$~s using
$f_r=10$~Hz.

\begin{table}[ht]
\caption{\label{tab:example1}  Phase shifts and operations assigned
to randomly generated numbers.}

\begin{ruledtabular}

\begin{tabular}{lllc}
$r$ in range & $\theta_1$ & $\theta_1$& Operation\\
$[0,p]$ & 0& 0 & $\openone$\\
$\left(p,p+\frac{(1-p)}{3}\right]$ & $\pi$ & 0& $\hat{Z}$\\
$\left(p+\frac{(1-p)}{3},p+\frac{2(1-p)}{3}\right]$ & 0 & $\pi$ & $\hat{X}$\\
$\left(p+\frac{2(1-p)}{3},1\right]$& $\pi$ & $\pi$ & $\hat{Y}$\\
\end{tabular}
\label{table_p_parameter}
\end{ruledtabular}
\end{table}

In the experiment, we prepared six different initial states,
eigenstates of operators $\{\hat{X}$, $\hat{Y}$, $\hat{Z}\}$ and projected them onto
the same set of states. We have thus registered photon counts for
all 36 combinations of prepared and projected states.

From the 36 measurements 
we calculated the PDO as well as the Choi-Jamio{\l}kowski matrix, using the
maximum likelihood estimation~\cite{Fiurasek2001}. From this
matrix, we calculated the robustness of the quantum memory \eqref{Eq_R_QM}
(marked by $\square$ in Fig.~\ref{fig_experimental_results}) and the
purity of the process.

From the same set of 36 measurements, we calculated the
assemblages $\rho_{\mathcal{E}}(a|x)$. In this case, for each of
the six preparation states, we obtained a full output state
tomography based on the six projection states, and the single-qubit
density matrices were estimated by maximum likelihood
\cite{Hradil1997}. Due to experimental imperfections, these
matrices slightly violate the no-signalling condition, which is
expressed as
\begin{equation}
\sum_a\rho_{\mathcal{E}}(a|x) = \sum_a \rho_{\mathcal{E}}(a|x')
~\forall~ x,x'.
\end{equation}
To solve this issue, we use semidefinite programming to find
assemblages $\{\tilde{\rho}_{\mathcal{E}}(a|x)\}$, that fulfil the
no-signaling condition, such that the sum of the fidelities of the
original unphysical assemblages violating the condition and the
physical ones $\{\tilde{\rho}_{\mathcal{E}}(a|x)\}$ is maximized.

The minimal fidelity across all parameters $v$ and all the
assemblages is 99.88\%. Using these newly found assemblages, we
calculated the robustness of temporal steering \eqref{Eq_TSR}
($\circ$ in Fig.~\ref{fig_experimental_results}).

For the calculation of the robustness of the non-NLB channel~\eqref{Eq_R_NLB},
we projected the process density matrix into eigenstates of $\{\hat{X}, \hat{Z}\}$ and $\{(\hat{X}+\hat{Z})/2,
(\hat{X}-\hat{Z})/2\}$ for the input and output qubits respectively
($\triangle$ in Fig.~\ref{fig_experimental_results}).

The details of the experimental results are presented in
Tables~\ref{tab:example1} and~\ref{tab:example2}.

\begin{table}[h]
\caption{\label{tab:example2} Table with experimental results}

\begin{ruledtabular}

\begin{tabular}{lcccc}
$v$ & $\text{R}_{\text{n-MR}}$ & $\text{R}_{\text{TS}}$ & $\text{R}_{\text{QM}}$ & Purity\\
0 & 0.000(0)& 0.000(0) & 0.000(0)& 0.255(1)\\
0.1 & 0.000(0)& 0.000(0) & 0.000(0)& 0.273(1)\\
0.2 & 0.000(0)& 0.000(0) & 0.000(0)& 0.290(2)\\
0.3 & 0.000(0)& 0.000(0) & 0.000(0)& 0.332(2)\\
0.35 & 0.000(0)& 0.000(0) & 0.000(4)& 0.338(2)\\
0.4 & 0.000(0)& 0.000(0) & 0.119(7)& 0.381(3)\\
0.5 & 0.000(0)& 0.000(0) & 0.237(7)& 0.434(3)\\
0.55 & 0.000(0)& 0.000(0) & 0.299(8)& 0.467(4)\\
0.6 & 0.000(0)& 0.008(2) & 0.382(6)& 0.514(3)\\
0.65 & 0.000(0)& 0.038(2) & 0.454(5)& 0.557(4)\\
0.7 & 0.000(0)& 0.073(2) & 0.538(5)& 0.612(4)\\
0.75 & 0.010(2)& 0.095(2) & 0.588(5)& 0.648(3)\\
0.8. & 0.050(2)& 0.131(2)& 0.675(5)& 0.712(4)\\
0.85 & 0.074(2) & 0.160(2)& 0.742(5) & 0.766(4)\\
0.9 & 0.094(2)& 0.194(1)& 0.820(3) & 0.832(3)\\
0.95 & 0.128(2) & 0.223(1) & 0.888(3) & 0.893(3)\\
1 & 0.161(1)& 0.258(1)& 0.972(1) &  0.972(1)\\
\end{tabular}
\end{ruledtabular}
\end{table}

\section{The negativity of a quantum channel as a memory measure}\label{App_f_function}
\red{
The negativity of a quantum channel $\mathcal{E}$~\cite{Holevo_Werner2001} is defined as
\begin{equation}\label{Eq:negativity of channel}
    \|\mathcal{E}\circ \mathcal{T}\|_{\lozenge}:=\sup_{\tilde{\rho}_{A_0,B_0}}\|\openone\otimes\mathcal{E}\circ\mathcal{T} (\tilde{\rho}_{A_0,B_0})\|_1,
\end{equation}
where $\mathcal{T}$ is the transpose under the computational basis, $\tilde{\rho}_{A_0,B_0}$ is a quantum state, and $\|X\|_{1(\lozenge)}$ is the trace (diamond) norm of operator $X$. 
Here, we also recall some properties of the negativity of quantum channels~\cite{Holevo_Werner2001}, which are used later: (i) If $\mathcal{E}$ is a CPTP map, then $\mathcal{T}\circ \mathcal{E} \circ \mathcal{T}$ is CPTP map too. Moreover, if $\II$ is a CP trace-nonincreasing map, so is $\mathcal{T}\circ \II \circ \mathcal{T}$. It can be proven, by noticing that $\mathcal{T}$ is trace invariant and $\mathcal{T}\circ X \circ \mathcal{T}$ preserves the property of CP when $X$ is CP.
(ii) $\mathcal{T}\circ \mathcal{T}=\openone$.}

\red{
Here, we recall that a particular free operation of quantum memory~\cite{Rosset2018} is
a pre-quantum instrument and the post-quantum channel with
classically shared randomness, namely,
\begin{equation}\label{Eq:free_memory}
\Lambda(\mathcal{E})=\sum_i D_i\circ \mathcal{E}\circ \II_i,
\end{equation}
where $\Lambda(\mathcal{E})$ is a free transformation acting on
the initial quantum channel $\mathcal{E}$.
Here, $D_i$ is a collection of quantum channels described by CPTP maps, and $\II_i$ is a quantum instrument satisfying CP trace-nonincreasing, which sums up to CPTP. It has been shown that any EB channel after operation is still EB. The map $\Lambda$ is a free
transformation since it transforms an EB channel to another EB
channel.}

\red{We now show that the negativity of the quantum channel satisfies monotonicity of a quantum memory with some assumptions.
We consider that the quantum instrument in the free operation of a quantum memory is constructed by a convex combination of quantum channels, namely $\mathcal{I}_i=p(i)D_i$. This assumption has also been widely studied in the channel discrimation problem~\cite{doi:10.1080/09500340008244034,Acin2001,Sacchi2005,Duan2009,Piani2009,Stefano2017,Pirandola2018,Pirandola2019}.}

\red{
We first note that property (i) in Appendix \ref{App_monotones} has been derived in Ref.~\cite{Holevo_Werner2001}. Therefore, when the channel is EB, the negativity of a quantum channel is zero. Here, we focus on showing that the negativity of a quantum channel satisfies the property (ii) in Appendix \ref{App_monotones} with an additional assumption.}

\red{Before we show the monotonicity, we first prove that the diamond norm of a quantum instrument consisting of a convex mixture of quantum channels satisfies:
\begin{lemma}\label{lemma:diamond norm of quantum instrument2}
If a quantum instrument is constructed by a convex mixture of quantum channels $\mathcal{I}_i= p(i)D_i$, we have
\begin{equation}
    \sum_i \|\mathcal{I}_i\|_{\lozenge}~= ~1.
\end{equation}
\end{lemma} 
\emph{Proof.---} 
Following the definition of the diamond norm, we have
\begin{equation}
    \begin{aligned}
    \sum_i \|\mathcal{I}_i\|_{\lozenge}&=\sum_i\|p(i)D_i\|_{\lozenge}\\
    &= \sum_i p(i)\|D_i\|_{\lozenge}\\
    &=\sum_i p(i)=1.
    \end{aligned}
\end{equation}
Here, we use the property of the absolute homogeneity $\|\alpha X\|=|\alpha|\|X\|$ with a real number $\alpha$ and any operator $X$, and the diamond norm of a quantum channel $D$ is always $\|D\|_{\lozenge}=1$. We note that this result also suggests that the triangle inequality $\|\sum_i\mathcal{I}_i\|_{\lozenge}~\leq~
    \sum_i \|\mathcal{I}_i\|_{\lozenge}$ is always saturated because $\sum_i\mathcal{I}_i$ is a CPTP map.} 

\red{
\begin{lemma}
The negativity of a quantum channel is decreasing under a quantum-memory free operation with an assumption.
\end{lemma}
\emph{Proof.---} We first assume that the supremum occurs when $\tilde{\rho}_{A_0,B_0}=\rho^*$.
Due to the Hermitian property, we define the
spectral decomposition of $\openone\otimes \mathcal{E}\circ \mathcal{I}_i\circ \mathcal{T}(\rho^*):=\omega^+_i -\omega^-_i$ with $\omega^\pm_i\geq 0$ for each $i$. We can write
\begin{widetext}
\begin{equation}
    \begin{aligned}
    \|\sum_{i}D_i\circ \mathcal{E}\circ \mathcal{I}_i\circ \mathcal{T}\|_{\lozenge}&=
    \|\openone\otimes \sum_{i}D_i\circ \mathcal{E}\circ \mathcal{I}_i\circ \mathcal{T}(\rho^*)\|_1 \\
    &\leq \sum_i\|\openone\otimes D_i\circ \mathcal{E}\circ \mathcal{I}_i\circ \mathcal{T}(\rho^*)\|_1 \\
    &=\sum_i \|\openone\otimes D_i(\omega^+_i -\omega^-_i)\|_1 \\
    &\leq\sum_i (\|\openone\otimes D_i (\omega^+_i)\|_1 + \|\openone\otimes D_i (\omega^-_i)\|_1) \\
    &=\sum_i(\text{Tr}[\openone\otimes D_i(\omega^+_i)] + \text{Tr}[\openone\otimes D_i(\omega^-_i)]) \\
    &=\sum_i(\text{Tr}[(\omega^+_i)] + \text{Tr}[(\omega^-_i)]) \\
    &=\sum_i\text{Tr}[(\omega^+_i+\omega^-_i)] \\
    &=\sum_i\|(\omega^+_i-\omega^-_i)\|_1 \\
    &= \sum_i\|\openone\otimes \mathcal{E}\circ \mathcal{I}_i\circ \mathcal{T} (\rho^*)\|_1\\
    &\leq  
    \sum_i\|\mathcal{E}\circ \mathcal{I}_i\circ \mathcal{T}\|_{\lozenge}
    \end{aligned}\label{Eq:proof_monotone_1}
\end{equation}
\end{widetext}
Here, we use, in order, the triangle inequality, spectral decomposition, triangle inequality, trace preserving property for all $D_i$, orthogonal support of $\omega^\pm_i$ for each $i$, and the definition of the diamond norm.}

\red{
If we now insert the fact that $\mathcal{T}\circ \mathcal{T}=\openone$, we can write
\begin{widetext}
\begin{equation}
    \begin{aligned}
     \sum_i\|\mathcal{E}\circ \mathcal{I}_i\circ \mathcal{T}\|_{\lozenge}&= \sum_i\|\mathcal{E}\circ \mathcal{T}\circ \mathcal{T} \circ \mathcal{I}_i \circ \mathcal{T}\circ \mathcal{T} \circ \mathcal{T}\|_{\lozenge}\\
     &= \sum_i\|\mathcal{E}\circ \mathcal{T}\circ \mathcal{T} \circ \mathcal{I}_i \circ \mathcal{T}\|_{\lozenge}\\
     &\leq \sum_i\|\mathcal{E}\circ \mathcal{T}\|_{\lozenge}\| \mathcal{T} \circ \mathcal{I}_i \circ \mathcal{T}\|_{\lozenge}\\
     &= \sum_i\|\mathcal{E}\circ \mathcal{T}\|_{\lozenge}\|  \mathcal{I}'_i \|_{\lozenge}\\
     &\leq\|\mathcal{E}\circ \mathcal{T}\|_{\lozenge}\\
    \end{aligned}\label{Eq:proof_monotone_2}
\end{equation}
\end{widetext}
The first inequality holds because the diamond norm is sub-multiplicative, namely $\|X\circ Y\|_{\lozenge}\leq\|X\|_{\lozenge}\|Y\|_{\lozenge}$ for any linear operators $X$ and $Y$.
The second inequality holds by the following reasons:
when we define $\mathcal{I}'_i=\mathcal{T}\circ \mathcal{I}_i\circ \mathcal{T}$, due to the property (i) below Eq.~\eqref{Eq:negativity of channel}, $\mathcal{I}'_i$ is a valid quantum instrument (CP trace-nonincreasing). It is easy to see that the assumption on the instrument still holds before and after transposing the quantum instrument. Then, we use
Lemma~\ref{lemma:diamond norm of quantum instrument2} to finish the proof. $\square$}

\red{
Finally, given a convex combination of quantum memories $\sum_i p(i)\mathcal{E}_i$, the negativity of a quantum channel satisfies
\begin{equation}
    \|\sum_i p(i)\mathcal{E}_i\circ \mathcal{T}\|_{\lozenge}\leq\sum_i p(i)\|\mathcal{E}_i\circ\mathcal{T}\|_{\lozenge}.
\end{equation}
This property holds because of the triangle inequality and absolute homogeneity.
With the above lemmas, we show that the negativity of a quantum channel is a convex quantum memory monotone under a particular choice of the free operation.}

\red{Now, we introduce the negativity-like measure of the temporal nonseparability.
Since a PDO is not necessarily positive semidefinite, it is convenient to quantify the degree of the temporal nonseparability by the negativity-like measure~\cite{Fitzsimons2015}:
\begin{equation}\label{EQ_f-function}
f=\frac{\|P_{\mathcal{E}}\|_1-1}{2}.
\end{equation}
It has been shown that the negativity-like measure of the temporal nonseparability is a lower bound of the negativity of a quantum channel~\cite{Pisarczyk2019}. Therefore, we can use the negativity-like measure of the temporal nonseparability to quantify quantum memory.}





%

\end{document}